\documentclass[epj,nopacs,draft]{svjour}

\def\bea{\begin{eqnarray}}
\def\eea{\end{eqnarray}}
\def\beq{\begin{equation}}
\def\eeq{\end{equation}}
\def\nono{\nonumber}

\def\rad{\langle r^2\rangle_s}
\def\deg{^\circ}

\usepackage{graphics}
\usepackage{epsf}
\begin{document}

\title{Analysis and interpretation of new low-energy $\pi\pi$ scattering data}
\author{S.~Descotes\inst{1} \and N.H.~Fuchs\inst{2} \and L.~Girlanda\inst{3}
\and J.~Stern\inst{4}}
%
\institute{Department of Physics and Astronomy,
University of Southampton, Southampton SO17 1BJ, U.K.\\
\email{sdg@hep.phys.soton.ac.uk} \and
Department of Physics, Purdue University,
    West Lafayette IN 47907, USA\\ \email{nhf@physics.purdue.edu} \and
    Dipartimento di Fisica, Universit\`a di Padova and
INFN, Via Marzolo 8, I-35131 Padova, Italy\\ \email{luca.girlanda@pd.infn.it}
\and
Groupe de Physique Th\'eorique, Institut de Physique Nucl\'eaire, F-91406
Orsay-Cedex, France\\ \email{stern@ipno.in2p3.fr} }
\date{}
%
\abstract{
The recently published E865 data on charged $K_{e4}$ decays
and $\pi\pi$ phases are reanalyzed to extract values of the two S-wave
scattering lengths, of the subthreshold parameters $\alpha$ and
$\beta$, of the low-energy constants $\bar{\ell}_3$ and $\bar{\ell}_4$
as well as of the main two-flavour order parameters: $\langle\bar
uu\rangle$ and $F_{\pi}$ in the limit $m_u = m_d = 0$ taken at the
physical value of the strange quark mass. Our analysis is exclusively
based on direct experimental information on $\pi\pi$ phases below 800
MeV and on the new solutions of the Roy equations by Ananthanarayan
{\it et al}. The result is compared with the theoretical prediction
relating $2a_0^0-5a_0^2$ and the scalar radius of the pion, which was
obtained in two-loop Chiral Perturbation Theory. A discrepancy at the
1-$\sigma$ level is found and commented upon. 
\PACS{
      {PACS-key}{discribing text of that key}   \and
      {PACS-key}{discribing text of that key}
     } 
} 
\titlerunning{Analysis of low-energy $\pi\pi$ scattering data}
\authorrunning{S.~Descotes, N.H.~Fuchs, L.~Girlanda and J.~Stern}

\maketitle

\section{Introduction}
\label{intro}

Recently, a new measurement of $K^+_{e4}$ decay and of the $\pi\pi$
phase shift difference $\delta^0_0 - \delta^1_1$ has been published
\cite{pislak} by the Brookhaven E865 collaboration. New low-energy
$\pi\pi$ scattering data constitute a rare event which has not
happened since the last Geneva-Saclay experiment 25 years ago~\cite{rosselet},
and the corresponding determination of the isoscalar S-wave scattering
length $a^0_0=0.26\pm 0.05$. The new experiment~\cite{pislak}
improves the statistics by more than a factor of $10$,
and the outcome for $a_0^0$ is not only more accurate, but also points
towards a smaller central value as expected by the
standard version of
Chiral Perturbation Theory ($\chi$PT)~\cite{GL1}. It is crucial to know
the two S-wave scattering lengths $a_0^0$ and $a_0^2$ as precisely as
possible and to avoid confusing their model-independent extraction
from the data on one hand with theoretical $\chi$PT-based predictions
on the other hand. This is the main purpose of the present paper. The
principal model-independent tools of our analysis are Roy
equations~\cite{Roy} and their recent solution by Ananthanarayan,
Colangelo, Gasser and Leutwyler (ACGL)~\cite{ACGL}. In a suitable
kinematical range, Roy equations represent a rigorous consequence of
general properties of the scattering amplitude: analyticity, unitarity,
crossing symmetry and asymptotic bounds.  The ACGL solution uniquely
fixes the three phases relevant at low energies,
$\delta_0^0(E),\delta_1^1(E),\delta_0^2(E)$ for $E<E_0=800$ MeV in
terms of the two scattering lengths $a_0^0$ and $a_0^2$, and
experimental data above $s_0 = E_0^2$. In this way, sufficiently
precise data on phase shifts for $2 M_\pi < E < E_0$ can be converted
into a model-independent determination of the two S-wave scattering
lengths.  Conversely, the knowledge of $a_0^0$ and $a_0^2$ allows one
to determine any other low-energy $\pi\pi$ observables which might be
needed for a theoretical interpretation of experimental results in
terms of the chiral structure of QCD vacuum. Unfortunately, the
combination $\delta_0^0 -\delta_1^1$ near threshold (which is measured
in $K_{e4}$ experiments) is not strongly sensitive to $a_0^2$, for a
given $a_0^0$, even if the former is allowed to vary over the whole Universal Band
(UB).  This can be seen explicitly, for instance, from the ACGL
solutions of Roy equations.  We will show (Sec.~\ref{sec:fits}) that a
model-independent and relatively accurate determination of both S-wave
scattering lengths is nevertheless possible, if the existing $K_{e4}$
data on $\delta_0^0 - \delta_1^1$ are combined with the older
production data by Hoogland {\it et al.}~\cite{hoogland} and by Losty
{\it et al.}~\cite{losty}, concerning the $I=2$ S-wave for $E<
800$~MeV. Our result reads $a_0^0 = 0.228 \pm 0.012$, $a_0^2 = -
0.0382 \pm 0.0038$.  Subsequently (Sec.~\ref{sec:kmsf}), Roy equations
will be used once more to convert the S-wave scattering lengths into
the determination of the subthreshold parameters $\alpha$ and $\beta$
(as well as $\lambda_1\ldots \lambda_4$) introduced in
Ref.~\cite{KMSF1} and under a different name ($b_1\ldots b_6$) in
Ref.~\cite{BCEG}.  The expansion of the parameters $\alpha$ and
$\beta$ in powers of quark mass converges more rapidly than in the
case of scattering lengths. The knowledge of subthreshold parameters
is therefore used to extract the low-energy constants (LEC's) $\bar \ell_3$
and $\bar \ell_4$, as well as the main two-flavour order parameters
(Secs.~\ref{sec:trunc}
and \ref{sec:l3l4}). All this analysis is based on existing solutions of Roy equations
as given by ACGL in Ref.~\cite{ACGL}. In parallel, we construct an
extended solution of Roy equations, incorporating the dependence on
the value of phase shifts at the matching point $E_0 = 800$ MeV.  This
allows the control of the propagation of errors arising from the
intermediate energy data, which was not possible using the published
ACGL solution of Ref.~\cite{ACGL}.  Finally, our model-independent
determination of scattering lengths and other low-energy parameters is
compared with the theoretical prediction of the tight correlation
between $2a_0^0-5a_0^2$ and the scalar radius of the pion
$\rad$~\cite{CGLPRL}. If the latter prediction is combined with the
E865 data alone \cite{pislak}, one obtains $a_0^0 = 0.218 \pm 0.013$,
$a_0^2 = - 0.0449 \pm 0.0033$ assuming the value of the scalar radius
$\rad = 0.61 \pm 0.04 \textrm{ fm}^2 $. Our simultaneous fit to the $K_{e4}$
\cite{pislak,rosselet} and to the high-energy $\pi^+\pi^+$-production data
\cite{hoogland,losty} suggests a discrepancy at the 1-$\sigma$ level
with the two-loop $\chi$PT prediction for $2a_0^0-5a_0^2$~\cite{CGLPRL}.
We point out that the treatment of symmetry breaking $O(p^6)$ corrections
in the scalar channel might be at the origin of this discrepancy
(Sec.~\ref{sec:scalar}).

\section{Theoretical motivation}

Before we proceed further it may be useful to summarize briefly the
main theoretical motivations which drive our interest in accurate
low-energy $\pi\pi$ scattering experiments, eventually completed by
independent experimental information of a different kind. In
particular it is worth explaining why a small difference in the values
of scattering lengths like the one mentioned in the introduction may
be relevant. The reader who is mostly interested in the analysis and
less in the interpretation of the new data can skip this section and
proceed directly to the following one. Nothing in our analysis depends
on theoretical ideas summarized here.

During the last two years our understanding of the pattern of chiral
symmetry breaking in QCD has considerably evolved
\cite{DGS,bachir,bachir2,DS,D,GST}, in particular concerning the
dynamical role of the number of light flavours.  $\pi\pi$ scattering
provides important information about the $SU(2)\times SU(2)$ chiral
structure of QCD vacuum in the limit of massless $u$ and $d$
quarks. The theoretical interpretation of this information, however,
remains incomplete unless one learns how to disentangle the influence
of the strange quark.  If the strange quark were massless, we deal
with $SU(3)\times SU(3)$ symmetry, which is the theoretician's
paradise \cite{leutlast}.  The reason is that all remaining quarks
($c$, $b$, $t$) are heavy compared to the QCD scale, and consequently
their influence on the chiral structure of the vacuum remains
tiny. For the same reason, we would reach another (a priori different)
paradise if the strange quark mass were sent to infinity. In these two
limiting cases, the question of the pattern of chiral symmetry
breaking reduces to the question of condensation of massless $\bar q q$
pairs in the vacuum; $\pi\pi$ scattering would by itself suffice to
detect and fully control this one-fermion loop effect. Unfortunately,
the deplorable fact that we live in neither of these paradises
\cite{leutlast} complicates the problem a bit. In the real world the
strange quark is considerably heavier than $u$ and $d$ quarks and yet
it is light compared to the QCD scale. Consequently, the vacuum gets
polluted by \emph{massive} virtual $\bar ss$ pairs. One may wonder how
this fact could affect the $SU(2)\times SU(2)$ chiral structure of the
vacuum, which is merely a matter of massless $u$ and $d$
quarks. Indeed, there would be no such effect in the large-$N_c$
limit, in which the transition:
\begin{equation}\label{OZI}
\bar s s  \longleftrightarrow  \bar u u  +  \bar d d
\end{equation}
is forbidden and, consequently, strange and non-strange virtual pairs
remain uncorrelated. Today we know that in the real world, the
OZI-rule violating transition Eq.~(\ref{OZI}) is in fact rather
important precisely in the vacuum channel \cite{DS} and that the
resulting symmetry breaking correlation $\langle \bar ss (\bar uu +
\bar dd)\rangle$ affects the $SU(2)\times SU(2)$ chiral structure of
the vacuum. We call this phenomenon which proceeds via at least two
fermion loops \emph{the induced quark condensate}~\cite{GST}.  It is
proportional to $m_s$, it enhances the effect of the genuine
condensate of massless quarks characteristic of the ideal world with
$SU(3)\times SU(3)$ symmetry, and it would persist even if the latter
would be absent. The situation can be summarized by the formula:
\begin{equation}\label{condensates}
\Sigma (2) = \Sigma (3) + m_s Z_{\rm{scalar}}\,,
\end{equation}
where $\Sigma(N_f)$ denotes the v.e.v. $-\langle\bar uu\rangle$ of the
lightest quark in the limit in which the first $N_f$ quarks become
massless, whereas the second term represents the induced condensate
with $Z_{\rm{scalar}}>0$ proportional to the amplitude of the
transition Eq.~(\ref{OZI}) in the scalar channel. Since $\Sigma(3)$ is
defined in the limit $m_u = m_d = m_s = 0$ and since there are no more
massive quarks left which would be sufficiently light to pollute the
vacuum, $\Sigma(3) = F^2_0 B_0$ (in the current $\chi$PT notation)
represents the true condensate of massless quarks of the $SU(3)\times
SU(3)$ symmetric world. On the other hand, $\Sigma(2)$,
defined at $m_u=m_d=0$ and the physical value of $m_s$, is detected in low
energy $\pi\pi$ scattering. Our original quest for the importance of quark 
condensation~\cite{FSS,KS} did not 
sufficiently emphasize the subtlety of the theoretical and experimental 
distinction between $\Sigma(2)$ and $\Sigma(3)$. Phenomenological 
indications in favour of a substantial OZI-rule violating transition,
Eq.~(\ref{OZI}), were not yet available and the large-$N_c$ wisdom 
$\Sigma(2) \sim \Sigma(3)$ dominated our thinking. Today, we have to answer 
the more precise question whether the suppression of the quark condensate --
expected in QCD due to the screening effect of multiflavour massless quark 
loops -- is already visible for $N_f=3$.

The important consequence of the new E865 data \cite{pislak} is that
$\Sigma(2)$ is sufficiently large to keep the two-flavour GOR ratio
$X(2)= (m_u+m_d) \Sigma(2) / F^2_{\pi} M^2_{\pi}$ close to one
\cite{CGL1} (see Sec.~\ref{sec:l3l4} for more details). However, we still do not
know whether the observed size of $\Sigma(2)$ reflects an important
contribution of the genuine condensate $\Sigma(3)$ or whether it is
(at least partially) due to the induced condensate and to the strange
quark mass, i.e. to the second term in Eq.~(\ref{condensates}). In the
latter case, the genuine condensate $\Sigma(3)$ could be rather small
and the corresponding three-flavour GOR ratio $X(3)= (m_u+m_d)
\Sigma(3) / F^2_{\pi} M^2_{\pi}$ significantly below one.  [Since
$X(2) \sim 1$, the suppression of $X(3)$ would imply an important
contribution of $m_s$ to $M^2_{\pi}$ through terms $m m_s , m m^2_s
\ldots $, where $m = (m_u + m_d)/2$.]  Moussallam \cite{bachir} has
proposed a rapidly convergent sum rule that allows an estimate of the
transition Eq.~(\ref{OZI}) and of the size of the induced
condensate. Detailed studies of this sum rule~\cite{bachir,bachir2,D}
are compatible with a suppression of the condensate by a factor $\sim
2$ when going from 2 to 3 flavours: $ X(3) \sim X(2)/2 $ indicating
that the two contributions to $\Sigma(2)$ on the right-hand side of
Eq.~(\ref{condensates}) should be of comparable size.  It is rather
difficult to make a precise quantitative statement here: the
evaluation of the Moussallam's sum rule involves a few steps that
include some unproved properties of the scalar form-factors and a
certain model-dependence (of a similar type to the one in the
phenomenological evaluation of the scalar radius of the
pion~\cite{DGL}). Hence, it is crucial to find an independent way of
disentangling the genuine and induced condensate contributions to
$\Sigma(2)$.

An analysis of $\pi\pi$ scattering exclusively based on $SU(2)$ $\times
SU(2)$ $\chi$PT can never separate the two components of $\Sigma(2)$
in Eq.~(\ref{condensates}), for the simple reason that it does not
involve any information on the actual size of $m_s$. On the other
hand, an $SU(3)\times SU(3)$ analysis of suitable $\pi\pi$ observables
supplemented by additional observables, such as $M_K$ or $F_K$,
directly sensitive to $m_s$, can considerably restrict the possible
values of three important parameters: \emph{i)} the three-flavour
GOR ratio $X(3)=2m\Sigma(3)/F^2_{\pi} M^2_{\pi}$, \emph{ii)} the quark
mass ratio $r=m_s/m$ and \emph{iii)} the pseudoscalar decay constant
$F_0 = F_{\pi}|_{m_u=m_d=m_s=0}$. An example of a relation of this
type is the strong correlation which exists between $r$ and $X(2)$
\cite{DGS,D}. Further restrictions can be obtained from the emerging
analysis of $K-\pi$ scattering \cite{Kpi}.

We are going to present this combined analysis in a separate
paper~\cite{next}. It will be shown in particular that in order to be
conclusive, one has to use extremely precise values of subthreshold
$\pi\pi$ parameters $\alpha$ and $\beta$ as the input of our analysis.
Typically, the actual precision driven by the new E865 data analyzed
in the following section will be just sufficient to reach conclusions
about the suppression of the genuine condensate $X(3)$ on the
1-$\sigma$ level. Notice that a more precise measurement of low-energy
$\pi\pi$ scattering is conceivable in more dedicated experiments which
are either ongoing~\cite{dirac} or planned~\cite{NA48}.

\section{The two S-wave scattering lengths}
\label{sec:fits}
\subsection{Extended solutions of Roy equations}

The E865 data can be analyzed using a parametric representation of the
solution of Roy equations for the $\pi\pi$ phase shifts. A set of
dispersion relations derived by Roy \cite{Roy} allows one to relate
the phase shifts $\delta_0^0$, $\delta_1^1$ and $\delta_0^2$ in the
region $4M_\pi^2 \leq s \leq s_0$ (with $\sqrt{s_0}\sim 0.8$ GeV) to
data at intermediate energies ($\sqrt{s_0} \leq \sqrt{s}\leq 2$ GeV) and to two
subtraction constants.  The latter can be identified with the two
S-wave scattering lengths $a_0^0$ and $a_0^2$.

The Roy equations, analyzed thoroughly in Ref.~\cite{ACGL}, yield
a boundary value problem. The solutions must interpolate between the 
phase shifts at the threshold, fixed by $a_0^0$ and $a_0^2$, and the 
three phases at the matching point $s_0$:
\begin{equation}
\theta_0=\delta_0^0(s_0),\qquad \theta_1=\delta_1^1(s_0),\qquad
\theta_2=\delta_0^2(s_0),
\end{equation}
determined from data above $s_0$.
As stated in Ref.~\cite{ACGL}, the behaviour of the phases above $s_0$ is
less important than the boundary values, because they only affect the slope
and the curvature of the solutions. With the experimental input encountered
in practice, the system of Roy equations admits a unique solution provided
that the matching point $s_0$ is carefully chosen
($0.78$~GeV $\leq \sqrt{s_0}\leq 0.86$~GeV).

Moreover, for given boundary conditions $(a_0^0,\theta_0,\theta_1,\theta_2)$,
arbitrary values of $a_0^2$ generate a strong cusp in the P-wave solution
at the matching point $s_0$. If we require the phases to be smooth, $a_0^2$
is determined as a function of $(a_0^0,\theta_0,\theta_1,\theta_2)$.
Since $\theta_{0,1,2}$ can only vary in their experimental range, 
this requirement leads to a correlation between $a_0^0$ and $a_0^2$, defining 
the so-called Universal Band (UB) in the $(a_0^0,a_0^2)$ plane.
Different choices for $\theta_{0,1,2}$ represent lines in 
the UB, $a_0^2=F(a_0^0)$. Inverting the relation between $a_0^2$ and 
$(a_0^0,\theta_0,\theta_1,\theta_2)$, we can consider $\theta_2$ as a function
of the other parameters. This means the solutions of the Roy equations
do depend eventually on $(a_0^0,a_0^2,\theta_0,\theta_1)$ only.

The data in the $I=0,1$ channels lead to:
\begin{equation}\label{eq:thetarange}
\theta_0=82.3\deg\pm   3.4\deg, \qquad \theta_1 = 108.9\deg\pm   2\deg.
\end{equation}
The authors of Ref.~\cite{ACGL} have provided explicit numerical
solutions of the Roy equations for $\theta_0=82.0\deg$ and
$\theta_1=108.9\deg$.  We have included in the parametrization of
Roy equations' solutions\footnote{ We have used the standard routines
of MINUIT for all the minimization and fitting procedures of this
paper.}, an explicit dependence on $\theta_0$ and $\theta_1$, 
generating solutions (with the same driving terms and experimental
input above the matching point $s_0$) for nine different sets
$(\theta_0,\theta_1)$ and a few tens of $(a_0^0,a_0^2)$ inside the UB.

Following Ref.~\cite{ACGL}, we parametrize our solutions, for energies
below 800~MeV, as:
\begin{eqnarray} \label{eq:deltaIl}
\tan \delta^I_\ell(s)
 &=&\sqrt{1-\frac{4M_\pi^2}{s}} q^{2\ell}
 \left\{A^I_\ell+B^I_\ell q^2+C^I_\ell q^4+D^I_\ell q^6\right\} \nonumber \\
 && \times
 \left(\frac{4M_\pi^2-s^I_\ell}{s-s^I_\ell}\right)\,.
\end{eqnarray}
The dependence on $a_0^0$ and $a_0^2$ of 
the Schenk parameters $X=A,B,C,D$ in Eq.~(\ref{eq:deltaIl}) is well approximated by:
\begin{eqnarray} \label{eq:paramACGL}
X^I_\ell&=&z_1+z_2 u+z_3 v+z_4 u^2+z_5 v^2 +z_6 uv \nonumber \\
&& +z_7 u^3+z_8 u^2 v+z_9 uv^2
  +z_{10} v^3\,,
\end{eqnarray}
where:
\begin{equation}
u=\frac{a_0^0}{p_0}-1\,,\quad
v=\frac{a_0^2}{p_2}-1\,,\quad
p_0=0.225\,, \quad p_2=-0.03706\,,
\end{equation}
while, for each coefficient $z_i$, the dependence on
the phase shifts at the matching point is parametrized by:
\begin{equation} \label{eq:paramz}
z_j=a_j+\delta\theta_0\, b_j +\delta\theta_1\, c_j\,,
\end{equation}
where
\begin{equation}
\delta\theta_0=\theta_0-82.3^\circ\,,\qquad 
\delta\theta_1=\theta_1-108.9^\circ\,.
\end{equation}
The parameters $s_0^0$, $s_1^1$ and $s_0^2$ are fixed by the boundary
conditions:
\begin{equation}
\delta_0^0(s_0)\equiv\theta_0\ ,\qquad \delta_1^1(s_0)\equiv\theta_1\ ,
\qquad \delta_0^2(s_0)\equiv\theta_2\ ,
\end{equation}
where $\theta_2(a_0^0,a_0^2,\theta_0,\theta_1)$ is parametrized 
following Eqs.~(\ref{eq:paramACGL}) and (\ref{eq:paramz}).
The coefficients $a_j$, $b_j$, $c_j$ are collected in App.~\ref{app:roy}.
This parametrization describes our solutions to better than $0.3\deg$
for the $I=0,2$, and $0.5\deg$ for the $I=1$ partial waves in the Universal band.  

By setting $\theta_0=82.0^\circ$ and $\theta_1=108.9^\circ$, 
we can compare with Ref.~\cite{ACGL}. We obtain slightly
different Schenk parameters for the so-called reference point 
$a_0^0=0.225$ and $a_0^2=-0.0371$, but the phase shifts are identical up 
to a few tenths of a degree. We obtain the same Universal Band, and 
only its lower half meets the consistency condition (Roy equations
fulfilled in their range of validity $2M_\pi\leq \sqrt{s}\leq \sqrt{s_1} = 1.15$ GeV). 
In the range of interest for $a_0^0$, the gap between the parametrization 
of Ref.~\cite{ACGL} and our Roy solutions is at most $0.3^\circ$ in the 
$I=0,2$ channels and $0.7^\circ$ in the $I=1$ channel (for $\sqrt{s}\sim 0.7$
GeV and much smaller near threshold in all the channels).

\subsection{Model-independent determination of $a_0^0$ and $a_0^2$}

The E865 data on $\delta_0^0 - \delta_1^1$ were analyzed in Ref.~\cite{pislak} in order 
to extract $a_0^0$. While two different (although compatible) results were quoted in
this reference for $a_0^0$, ($a_0^0=0.228 \pm 0.012 \pm
0.004^{+0.006}_{-0.012}$ and $a_0^0 = 0.216 \pm 0.013 \pm 0.004 \pm 0.005$),
no results were given for $a_0^2$, which is harder to pin down from $K_{e4}$
data alone (see Sec.~\ref{intro}). An unconstrained fit of E865 data, 
using the Roy equations of Ref.~\cite{ACGL}, leads to a 
rather inaccurate result for
the $I=2$ scattering length: $a_0^0=0.237\pm 0.033$ and $a_0^2=-0.0305 \pm 0.0226$
($\chi^2=5.44/5$~d.o.f and the correlation coefficient is 0.96).

In order to extract both scattering lengths additional information has to
be provided. The first constraint arises from the necessary consistency
of the Roy solutions with the $I=2$ data above the matching point.
This forces the S-wave scattering lengths to lie within the 
so-called Universal Band. Unfortunately, this model-independent
constraint is rather weak. 

We make use of additional information by fitting a broader set of data
below 800 MeV, namely Rosselet and E865 sets for $I=0,1$
\cite{rosselet,pislak} and Hoogland (sol. A) and Losty sets for $I=2$
\cite{hoogland,losty}.  Notice that a similar fit has been considered
in Refs.~\cite{ACGL,ochs} but without the E865 data (cf. Figs.~11 and
12 in Ref.~\cite{ACGL}).

We first perform a 
fit using the solutions of the Roy equations of Ref.~\cite{ACGL}. 
The $\chi^2$ is defined as:
\begin{eqnarray}
&&\chi^2_{\rm global}(a_0^0,a_0^2)=
\sum_{j=1}^9 \left(
\frac{(\delta_0^2)^{\rm ACGL}(s_j^{\rm exp})
     -(\delta_0^2)_j^{\rm exp}}{\sigma_j^{\rm exp}}\right)^2 \nonumber \\
&& + \sum_{i=1}^{11} \left(
\frac{[\delta_0^0-\delta_1^1]^{\rm ACGL}(s_i^{\rm exp})
     -[\delta_0^0-\delta_1^1]_i^{\rm exp})}{\sigma_i^{\rm exp}}\right)^2 ,
\end{eqnarray}
where $[\delta^I_\ell]^{\rm ACGL}(a_0^0,a_0^2,s)$ is the
parametrization of Roy solutions proposed in Ref.~\cite{ACGL}. $i$ and
$j$ are the indices of the experimental points for $I=(0,1)$ and $I=2$
respectively. This fitting procedure, referred to as ``global'',
yields:
\begin{equation}
\textrm{Global:}\quad a_0^0=0.228\pm   0.012,\quad a_0^2=-0.0382\pm   0.0038.
\end{equation}
with $\chi^2_{\rm min}=16.45 / 18$ d.o.f and a correlation coefficient of
0.788.
Including data on the P-wave below 800~MeV reported by Protopopescu {\em et
al.} \cite{proto} does not change the
result of the fit, yielding\footnote{Specifically we have used the
energy-independent solution, which has larger error bars. The P-wave
production data at low energy have been often criticized, most recently in
Ref.~\cite{ACGL}. Therefore we prefer not to include them in the analysis,
but instead use our results to predict the P-wave at low energy.}
$\chi^2_{\rm min}=23.1 / 28$~d.o.f. and $a_0^0=0.228 \pm 0.012$,
$a_0^2=-0.0392 \pm 0.0038$.

A second fitting procedure can be followed, in which 
we use our solutions of the Roy equations to include the dependence
on the phase shifts at the matching point $(\theta_0,\theta_1)$.
The $\chi^2$ is then defined as:
\begin{eqnarray}
&&\chi^2_{\rm extended}(a_0^0,a_0^2,\theta_0,\theta_1)=
\sum_{j=1}^{9} \left(
\frac{(\delta_0^2)^{\rm ext}(s_j^{\rm exp})
     -(\delta_0^2)_j^{\rm exp}}{\sigma_j^{\rm exp}}\right)^2
\nonumber \\
&& + \sum_{i=1}^{11} \left(
\frac{[\delta_0^0-\delta_1^1]^{\rm ext}(s_i^{\rm exp})
     -[\delta_0^0-\delta_1^1]_i^{\rm exp})}{\sigma_i^{\rm exp}}\right)^2
\nonumber \\
&&  +\left(\frac{\theta_0-82.3\deg}{3.4\deg}\right)^2
  +\left(\frac{\theta_1-108.9\deg}{2\deg}\right)^2, 
\end{eqnarray}
where $[\delta^I_\ell]^{\rm ext}(a_0^0,a_0^2,\theta_0,\theta_1,s)$ 
is our extended parametrization of Roy solutions.
This fit, called ``extended'', leads to the same S-wave scattering lengths as  the ``global'' fit:
\begin{equation}
\begin{array}{lll}
\textrm{Extended:} & a_0^0=0.228\pm   0.013,  & a_0^2=-0.0380\pm   0.0044,\\
& \theta_0=82.1\deg\pm   3.3\deg,  & \theta_1=108.9\deg\pm   2.0\deg 
\end{array}
\end{equation}
with $\chi^2_{\rm min}=16.48/18$ d.o.f and the correlation matrix:
\begin{equation}
\begin{array}{c|cccc}
& a_0^0 & a_0^2 & \theta_0 & \theta_1\\
\hline
a_0^0 & 1.000 & 0.799  & -0.319 & -0.004 \\
a_0^2 & -     & 1.000  & -0.271 & 0.029  \\
\theta_0 & - & - & 1.000 & 0.000 \\
\theta_1 & - & - & - & 1.000
\end{array}
\end{equation}

\begin{figure*}[ht]
\epsfxsize16cm
\centerline{\epsffile{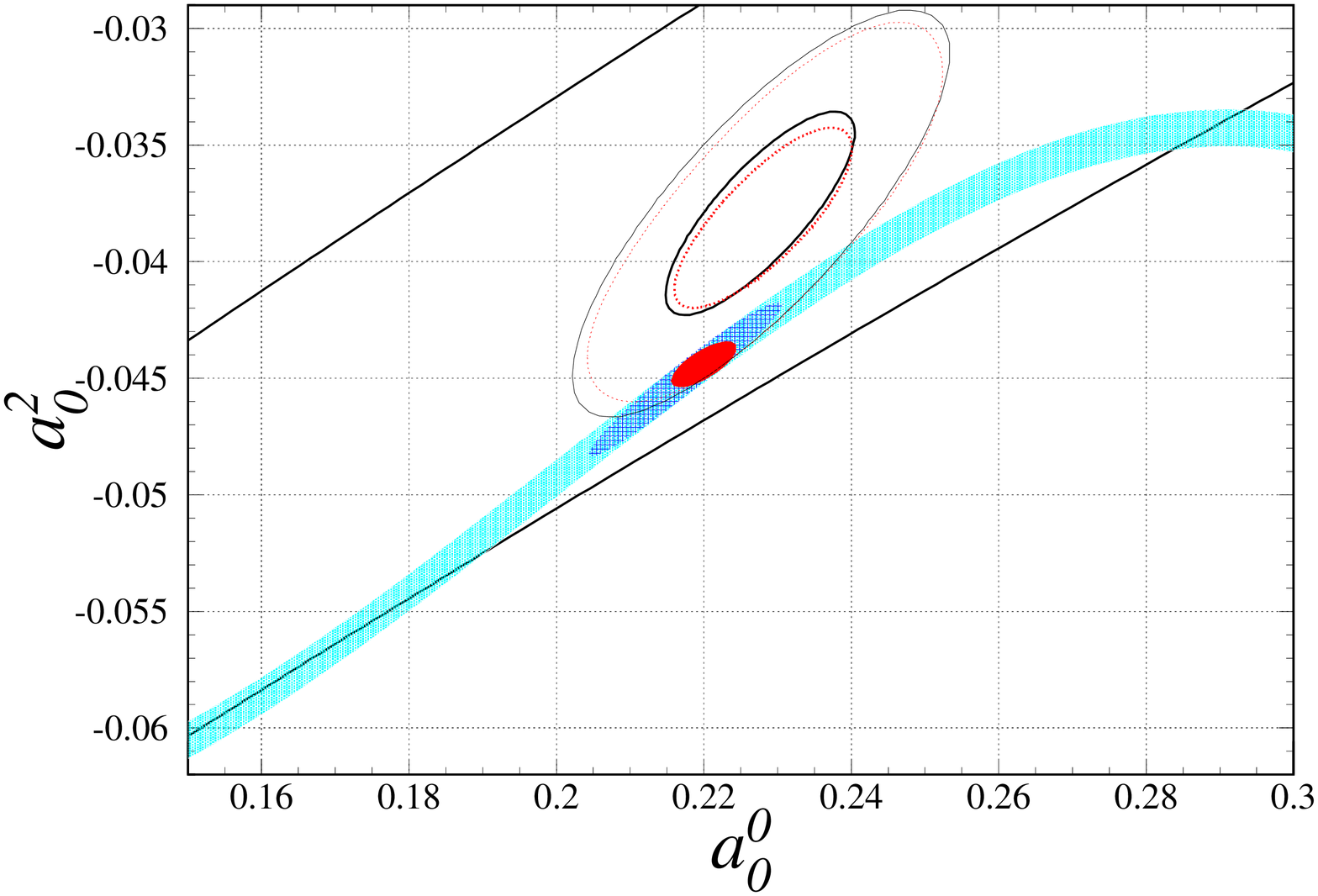}}
\caption{Fit results using either the Roy solutions of
Ref.~\cite{ACGL} (red, dotted ellipse -- ``global'') or our
parameterization of the Roy solutions (black, solid ellipse --
``extended''). In each case, the thicker lines indicate the 1-$\sigma$
ellipse, and the thinner ones the 2-$\sigma$ contour. The Universal
Band (delimited by the two straight lines) is drawn according to Ref.~\cite{ACGL}, and the
narrow strip (shaded region, cyan) related to the scalar radius of the
pion is taken from Ref.~\cite{CGLPRL}.  We have indicated the result
of the fit ``scalar'' (blue, hatched ellipse) and the S$\chi$PT
prediction of Ref.~\cite{CGLNPB} (red, filled small ellipse).}
\label{fig:fitres}
\end{figure*}

The results of these analyses are shown in Fig.~\ref{fig:fitres}, where
we have indicated the 1- and 2-$\sigma$ contours for both
determinations. These contours are defined respectively as\footnote{
We recall that in the case of the simultaneous determination of two
variables, this definition of the contours correspond to 39\% and 86\%
confidence level.}
$\chi^2=\chi^2_{\rm min}+1$ and $\chi^2=\chi^2_{\rm min}+4$.
We see that our fitting result lies slightly below the center of the UB,
where the consistency condition for Roy equations is met.

A comment is in order here concerning our use of the ACM(A) $I=2$ phase shifts
by Hoogland {\it et al.} These were extracted following the method A,
which is a conventional Chew-Low extrapolation to the pion pole of the measured
$t$-channel ($m=0$) helicity moments~\cite{hoogland} (the beam momentum was
14 GeV/$c$). A similar method was used by Losty {\it et al.}~\cite{losty}.
A second method (B) is presented in Ref.~\cite{hoogland}, which is based
on an overall fit of an (absorption) model for the amplitude to all
non-negligible $s$-channel helicity moments ($m=0$ and $m=1$). The method B
involves extra assumptions and parameters, some of which exhibit unexpectedly
rapid energy variations. No $\chi^2$ is given in Ref.~\cite{hoogland} (in a
preliminary analysis~\cite{hoogland2}, based on a third of the data, a poor
$\chi^2$ was reported for method B).

We have tried to use in our fit solution B of Ref.~\cite{hoogland}
instead of solution $A$. Due to the small error bars of the former, we did
not succeed in obtaining a consistent description of both ACM(B) and
E865-$K_{e4}$ data within the ACGL solutions of Roy equations~\cite{ACGL}.
The minimum has $\chi^2=68/18$ d.o.f. and is situated far outside the
Universal Band. Such a fit has little meaning, since the ACGL solutions are valid
exclusively inside the Universal Band. Solution ACM(A) is free from such
difficulties.

It has been suggested in Ref.~\cite{ACGL} that the difference between the phase
shifts ACM(A) and (B) indicates sizeable systematic errors, and that
the errors associated with ACM(A) solution should consequently be enlarged.
It is not obvious to us that method B yields a correct estimate of the
systematic errors inherent to method A -- especially since the two methods
do not use the same sample of data. We find it nevertheless useful to show
in App.~\ref{sec:enlarged} and in Fig.~\ref{fig:enlarged} how our results
would be modified if the errors in ACM(A) phase shifts were increased
according to the prescription advocated in Ref.~\cite{ACGL}. Let us mention
here that these modifications barely affect the main conclusion.

\subsection{Comparison with the $\chi$PT prediction for $2a_0^0-5a_0^2$}

In the theoretical prediction of $a_0^0$ and $a_0^2$ based on standard
$\chi$PT including $O(p^6)$ accuracy \cite{BCEG,CGLPRL,CGL1,CGLNPB},
one may distinguish two steps. The first step concerns the relation
between the combination $2a_0^0-5a_0^2$ and the scalar radius of the
pion $\rad$ \cite{CGLPRL,CGLNPB}.  This step is practically
independent of the badly known $O(p^4)$ constant $\bar{\ell}_3$ but it
requires an independent phenomenological determination of $\rad$ and
it is rather sensitive to the two-loop corrections (a more detailed
discussion of this theoretical prediction can be found in
Sec.~\ref{sec:scalar}).  If
one takes the value $\rad = 0.61 \pm 0.04 \textrm{\ fm}^2$, the
prediction amounts to a narrow strip in the $a_0^0 - a_0^2$ plane,
given in Ref.~\cite{CGLPRL} and reproduced in Fig.~\ref{fig:fitres}:
\begin{equation}\label{eq:corrscal}
 a_0^2=G(a_0^0)\pm   .0008,\\
\end{equation}
where the function $G(a_0^0)$ may be parametrized as
\begin{eqnarray} \label{eq:constscal}
G(a_0^0)&=&-.0444 +.236 (a_0^0-.22) -.61 (a_0^0-.22)^2 \nonumber \\
&& -9.9(a_0^0-.22)^3\,,
\end{eqnarray}
and the error bar  is estimated  within the theoretical
framework defined in Refs.~\cite{CGLPRL,CGLNPB} (see Sec.~\ref{sec:scalar}
of the present paper). The second step of the prediction procedure then
consists in locating the actual position inside the narrow strip
Eq.~(\ref{eq:corrscal}) and it involves, among other things, an estimate of the
constant $\bar\ell_3$.

The analysis performed in the previous subsection ma\-kes use only of
the E865 and Rosselet data on $\delta_0^0 - \delta_1^1$, the Hoogland
and Losty data on $\delta_0^2$ together with the solution of Roy
equations, and does not use $\chi$PT or a particular value of
$\rad$. It provides thus a sensitive experimental test of the
theoretical prediction represented by the CGL correlation
Eqs.~(\ref{eq:corrscal}) and (\ref{eq:constscal}). It is seen from
Fig.~\ref{fig:fitres} that the 1-$\sigma$ ellipses resulting from both
fits ``global'' and ``extended'' are situated outside the CGL narrow
strip despite the fact that they are entirely contained within the
Universal Band required by the consistency of Roy equations
solution. On the other hand, the 2-$\sigma$ contours intersect the
narrow strip. We thus conclude that there is a marginal (1-$\sigma$
deviation) discrepancy between the theoretical prediction
Eq.~(\ref{eq:constscal}) and the result of the ``global'' and
``extended'' fits.  This discrepancy will be further commented on in
Sec.~\ref{sec:scalar}.

                In order to make the comparison more quantitative, we can
 perform a fit to the E865 data alone, imposing
by hand the correlation described by the narrow strip. A similar fit  has
been performed in Ref.~\cite{CGLPRL}, leading
to the central value $a_0^0=0.218$ (no uncertainty was indicated 
in that reference). Our fitting procedure is  defined by:
\begin{eqnarray}
&&\chi^2_{\rm scalar}(a_0^0,a_0^2) =
\left(\frac{a_0^2-G(a_0^0)}{.0008}\right)^2 \nonumber \\
&& + \sum_{k=1}^6 \left(
\frac{[\delta_0^0-\delta_1^1]^{\rm ACGL}(s_k^{\rm exp})
     -[\delta_0^0-\delta_1^1]_k^{\rm exp}}{\sigma_k^{\rm exp}}\right)^2 ,
\end{eqnarray}
where $k$ runs only through the E865 points.  We have obtained:
\begin{equation}
\textrm{Scalar:}\quad a_0^0=0.218\pm   0.013,\qquad a_0^2=-0.0449\pm   0.0033.
\end{equation}
with $\chi^2_{\rm min}=5.89/5$ d.o.f., compatible with the results of
E865 \cite{pislak} for $a_0^0$, and a correlation coefficient of 0.972. The corresponding 1-$\sigma$ contour
is indicated in Fig.~\ref{fig:fitres}. We refer to this fit and to the
corresponding 1-$\sigma$ ellipse as ``scalar'' to be compared with the
``global'' and ``extended'' fits. The meaning of $\chi^2$ and of the
standard deviation in the ``scalar'' fit should be taken with caution:
The error bar 0.0008 arises from uncertainties in the experimental
input, while the theoretical errors inherent in the estimate of
$O(p^6)$ corrections are more difficult to quantify.  On the other
hand, the fits ``global'' and ``extended'' are fully based on
experimental data and corresponding errors.

Finally, we would like to briefly comment on the Olsson sum rule for $2
a_0^0 - 5 a_0^2$, as discussed in Ref.~\cite{ACGL}. This sum rule converges
slowly and demands good control
over the asymptotic contribution, which is hard to obtain outside specific
models. According to the model used for this purpose in Ref.~\cite{ACGL}, the
asymptotic contribution to $2 a_0^0 - 5 a_0^2$ is $O_{\mathrm{as}}=0.102\pm
0.017$. Even with such small error bar, the final result shown in Eq.~(11.2)
of Ref.~\cite{ACGL} is consistent with our global fit, which leads to $(2
a_0^0 - 5 a_0^2)_{\mathrm{global}} = 0.647 \pm 0.015$. If the actual error
bar in $O_{\mathrm{as}}$ is bigger, the impact of the Olsson sum rule on our
fit becomes completely negligible.

\section{Subthreshold parameters}\label{sec:kmsf}

In the low-energy domain the $\pi\pi$ amplitude is strongly
constrained by chiral symmetry, crossing and unitarity. As was first
shown in Ref.~\cite{FSS}, the amplitude depends on only six parameters
up to and including terms of order $(p/\Lambda_H)^6$ in the low-energy
expansion.  In Ref.~\cite{KMSF1}, the amplitude was written as:
\begin{eqnarray}
&& A_{\mathrm{KMSF}}(s|t,u)=A^{\rm cut}(s|t,u)+A^{\rm pol}(s|t,u)\\
&& A^{\rm pol}(s|t,u)=
  \frac{\beta}{F_\pi^2}\left(s-\frac{4M_\pi^2}{3}\right)
  +\frac{\alpha}{F_\pi^2}\frac{M_\pi^2}{3} \nonumber \\
&& +\frac{\lambda_1}{F_\pi^4} (s-2M_\pi^2)^2
   +\frac{\lambda_2}{F_\pi^4} \left[(t-2M_\pi^2)^2+(u-2M_\pi^2)^2\right]\nono\\
&& +\frac{\lambda_3}{F_\pi^6} (s-2M_\pi^2)^3
   +\frac{\lambda_4}{F_\pi^6} \left[(t-2M_\pi^2)^3+(u-2M_\pi^2)^3\right]
 \,, \nonumber \\
&& 
\end{eqnarray}
where $A^{\rm cut}$ is a known function of the Mandelstam variables
$s,t,u$ that collects the unitarity cuts of the amplitude and
explicitly depends on $\alpha,\beta,\lambda_1,\lambda_2$.

This amplitude was constructed as the general solution of unitarity,
analyticity and crossing symmetry up to and including $O(p^6)$.  The
six parameters $\alpha,\beta,\lambda_1,\ldots,\lambda_4$ correspond to
an expansion of the amplitude in the central region of the Mandelstam
triangle, and are therefore called ``subthreshold parameters''.  A
complete calculation in the framework of Standard $\chi$PT \cite{BCEG}
confirmed this result, allowing in addition to relate the six
parameters to the quark masses and LEC's of the standard chiral
Lagrangian.  This last step is crucial to translate the experimental
information into knowledge of the LEC's, which parametrize the chiral
structure of the vacuum of QCD. The six parameters introduced in
Ref.~\cite{BCEG}, $\bar b_1,\ldots ,\bar b_6$, are dimensionless
combinations of LEC's in one-to-one (linear) correspondence with
$\alpha,\beta,\lambda_1,\ldots ,\lambda_4$ of Ref.~\cite{KMSF1}, or
with $c_1,\ldots ,c_6$, subsequently introduced in
Ref.~\cite{CGL1}. Different choices for the set of six subthreshold
parameters correspond to different parametrizations of solutions of
unitarity, analyticity and crossing symmetry constraints, which are
equivalent up to $O(p^6)$ and only differ at $O(p^8)$.

On the other hand, the Roy equations allow one to determine the
low-energy amplitude in terms of only two subtraction constants,
identified with the two scalar scattering lengths.  It is therefore
possible to match the two amplitudes in their common domain of
validity, in order to determine, through the experimental
determination of the scattering lengths, the six subthreshold
parameters. Such a program was already advocated in Ref.~\cite{KMSF2},
leading to rapidly convergent sum rules for the parameters
$\lambda_1,\ldots,\lambda_4$.  A similar matching procedure, using new
solutions of the Roy equations, has been carried out in
Ref.~\cite{CGLNPB} (subtracting the dispersion integrals at $s=0$).
Let us briefly outline the various steps, using the notation and
results of this last reference.

Starting with particular values of $a_0^0$ and $a_0^2$, we can use the
solutions of the Roy equations to compute the low-energy moments
$J^I_n$. In conjunction with the background moments $I^I_n$ and $H$,
estimated in Ref.~\cite{ACGL}, we compute the phenomenological moments
$\bar{I}^I_n$ and their linear combinations denoted
$\bar{p}_{i=1\ldots 6}$ and defined in Eq.~(3.5) of
Ref.~\cite{CGLNPB}. Matching the phenomenological and the chiral
representations of the amplitude connects the phenomenological
parameters $\bar{p}_{i=1\ldots 6}$ and the chiral ones $c_{i=1\ldots
6}$ [see Eq.~(4.2) of the same reference]. App.~A and B of
Ref.~\cite{CGLNPB} can then be used to translate the chiral parameters
$c_{i=1\ldots 6}$ into the parameters $\bar{b}_{i=1\ldots 6}$ defined
in Ref.~\cite{BCEG}, and finally into $\alpha$, $\beta$ and
$\lambda_{i=1\ldots 4}$.

We can repeat this procedure for each set of $(a_0^0,a_0^2)$, as
determined from the ``global'' and ``scalar'' fits, or
$(a_0^0,a_0^2,$ $\theta_0,\theta_1)$, as determined from the ``extended''
fit, described in the previous section.  In order to take full account
of the theoretical and experimental correlations among the six
parameters, we proceed in the following way: we generate random sets
of $(a_0^0,a_0^2)$ or $(a_0^0,a_0^2,\theta_0,\theta_1)$, distributed
according to the 2- or 4-dimensional gaussian obtained from the
covariance matrix of the fit.  We then fit the resulting distributions
for the subthreshold parameters by gaussians, leading\footnote{ The
phenomenological moments $J^I_n$ are integrals of the $I=0,1,2$ phase
shifts from threshold to $\sqrt{s_2}=2$ GeV. The solutions of the Roy
equations are used for $s\leq s_0$, and experimental input is used
above the $K\bar{K}$ threshold.  An interpolation is necessary in the
intermediate region $[s_0,4M_K^2]$.  We have observed a weak
sensitivity of $\lambda_1$ and $\lambda_2$ on the interpolation
prescription. On the other hand, the values of $\alpha$ and $\beta$
are independent of this procedure.} to Table~\ref{tab:KMSF}.
$\rho_{\alpha\beta}$
denotes the correlation coefficient between $\alpha$ and $\beta$.

\begin{table}
\caption{Subthreshold parameters for the three different fits considered in this paper.
See the text for a discussion of the error bars.}\label{tab:KMSF}
\begin{tabular}{|c|c|c|c|}
\hline
& {Global}
& {Extended}
& {Scalar}
\\
\hline
 $\bar{b}_1$ & -2.08 $\pm$  6.12
             & -1.51  $\pm$  7.01  
             & -13.06 $\pm$  5.31\\
 $\bar{b}_2$ & 9.35  $\pm $  1.43
             & 8.93  $\pm$  1.62 
             & 11.51 $\pm$  0.62 \\ 
 $\bar{b}_3$ & -0.38 $\pm$   0.03
             & -0.36  $\pm$  0.07
             & -0.33  $\pm$  0.01\\ 
 $\bar{b}_4$ &  0.716 $\pm$   0.008
             & 0.710  $\pm$  0.010  
             & 0.727  $\pm$  0.005\\ 
 $\bar{b}_5$ &  3.21  $\pm$  0.25
             & 3.21  $\pm$  0.44 
             & 3.53  $\pm$  0.15\\ 
 $\bar{b}_6$ &  2.23  $\pm$  0.07
             & 2.20  $\pm$  0.08 
             & 2.34  $\pm$  0.03\\
\hline
 $\alpha$    &  1.381 $\pm$  0.242
             &  1.384 $\pm$  0.267
             &  1.034 $\pm$  0.248\\
 $\beta$     &  1.081 $\pm$  0.023
             &  1.077 $\pm$  0.025
             &  1.116 $\pm$  0.010\\
 $\rho_{\alpha\beta}$ 
        & -0.14 
        & -0.23
	& 0.53\\
 $\lambda_1 \cdot 10^3$ 
             & -4.40  $\pm$   0.28
             & -4.18  $\pm$   0.63
             & -3.97  $\pm$   0.12\\
 $\lambda_2 \cdot 10^3$ 
             &  9.04  $\pm$   0.10 
             &  8.96  $\pm$   0.12
             &  9.17  $\pm$   0.06\\ 
 $\lambda_3 \cdot 10^4$ 
             &  2.21  $\pm$   0.10
             &  2.22  $\pm$   0.16
	     &  2.32  $\pm$   0.06\\  
 $\lambda_4 \cdot 10^4 $
             & -1.40  $\pm$   0.04
             & -1.38  $\pm$   0.04
             & -1.46  $\pm$   0.02\\  
\hline
\end{tabular}
\end{table}

\begin{figure}[ht]
\epsfxsize10cm
\centerline{\epsffile{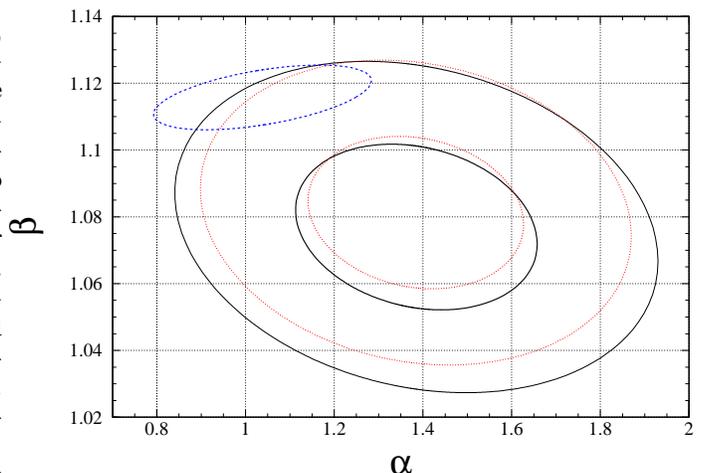}}
\caption{Correlation between $\alpha$ and $\beta$ from the fit ``scalar'' (blue, dashed ellipse) and our fits (black, solid for the
``extended''  and red, dotted for the ``global''). 
The thicker lines correspond to the 1-$\sigma$ ellipses, whereas the thinner
ones indicate the 2-$\sigma$ ellipses (not shown for the ``scalar'' fit).} 
\label{fig:abcorr}
\end{figure}

The slightly larger error bars of the ``extended'' fit, compared to
the ones of the ``global'' fit, reflect the influence of the
uncertainties in $\theta_0$ and $\theta_1$, which in the ``global''
fit are not explicitly taken into account. The differences in the
central values between these two fits (although compatible within the
errors) may be ascribed to the fact that the ``extended''
parametrization is not as accurate as the ACGL one, due to the fact
that the former has to account for the dependence on two more
variables.  The column referring to the ``scalar'' fit should be
understood as originating from a mixture of E865 data and
$\chi$PT-based theoretical predictions that rely on assumptions about
the size of $O(p^6)$ counterterms (see Sec.~\ref{sec:scalar}). For this
reason the associated errors should not be interpreted in the strict
statistical sense.  The corresponding 1- and 2-$\sigma$ ellipses in
the $(\alpha - \beta)$ plane are drawn in Fig.~\ref{fig:abcorr}.

It is worth stressing that a rather small difference between $a_0^0$
and $a_0^2$ resulting from the scalar fit, on the one hand, and from
the global and extended fits on the other hand, results in a more
pronounced difference in the corresponding values of the subthreshold
parameters $\alpha, \beta$.  Whereas the scalar fit (and the CGL
prediction) is characterized by values of $\alpha$ close to (or
smaller than) 1 and $\beta$ well above 1.10, the global and extended
fits lead to central values of $\alpha \sim 1.4$ and relatively
smaller values of $\beta$.  It will be shown elsewhere \cite{next}
that this qualitative difference finds its interpretation within the
three-flavour analysis of $\pi\pi$ scattering together with other
observables.

\section{$N_f = 2$ Mass and Decay constant Identities}
\label{sec:trunc}
In order to investigate the consequences of the results obtained so
far for the parameters of the effective Lagrangian, we start with the
Ward identity satisfied by the two-point function of the divergence of
the axial current $\bar u\gamma_{\mu}\gamma_{5} d$ and of its
conjugate at zero momentum transfer. We isolate all LO (linear) and
all NLO (quadratic) contributions in the quark mass $m=(m_u + m_d)/2$,
and collect all $O(m^3)$ and higher order terms into the NNLO
remainder $\delta$:
\begin{equation}\label{mass}
 F^2_\pi M^2_\pi  = F^2 M^2 + \frac{M^4}{32\pi^2}(4 \bar{\ell}_4 - \bar{\ell}_3) +\\
F^2_\pi M^2_\pi \delta . 
\end{equation}
Here
\begin{equation}\label{definitions}
M^2 = 2 m B,\qquad B= \Sigma(2)/F^2
\end{equation}
are defined in the $SU(2)\times SU(2)$ chiral limit, keeping the
strange quark mass at its physical value:
\begin{eqnarray}\label{limit}
F &=& \lim_{m_u,m_d\to 0}F_{\pi}|_{m_s ={\rm physical}}\,,\\
\label{condensate}
\Sigma(2) &=& - \lim_{m_u,m_d\to 0} \langle \bar uu \rangle|_{m_s =
  {\rm{physical}}} \,. 
\end{eqnarray}
A similar identity holds for the two-point function of axial minus
vector currents, giving
\begin{equation}\label{decayconstant}
F^2_\pi = F^2 + \frac{M^2}{8\pi^2}\bar{\ell}_4 + F^2_\pi \varepsilon.
\end{equation}
On the right hand side, all LO and NLO contributions are again
explicitly shown, and all higher orders - $O(m^2)$ and higher - are
included in the NNLO remainder $\varepsilon$. $\bar{\ell}_3$ and $\bar
\ell_4$ are the standard $SU(2)\times SU(2)$ LEC's, which are scale-independent
and exhibit logarithmic singularity in the chiral limit. They can be
defined non-perturbatively, via the low-energy behaviour of the
two-point functions that enter the Ward identities considered above.
It may be useful to consider the NNLO remainders $\delta$ and
$\varepsilon$ in Eqs.~(\ref{mass}) and (\ref{decayconstant}) as known,
and to treat the mass and decay constant identities exactly, not
performing any expansion. This avoids the use of perturbation theory
when eliminating the order parameters $M^2$ and $F^2$ in favour of
observable quantities. We will see later that, despite the fact that
$\alpha - 1$ need not be particularly small, this non-perturbative
precaution is not absolutely necessary in the $N_f=2$ case. It will
however, be fully justified in the case of three light flavours.
In any case, the above method -- which does not coincide either with the
Standard or with the Generalized $\chi$PT -- is completely
meaningful no matter how large $\bar{\ell}_3$ or how small the
condensate might be.

The fundamental order parameters, in appropriate units, the condensate
and the decay constant defined as:
\begin{equation}\label{order} 
X(2) = \frac{2m \Sigma(2)}{F^2_\pi M^2_\pi}\,,\quad 
Y(2) = \frac{2m B}{ M^2_\pi} \,,\quad 
Z(2) = \frac{F^2}{F^2_\pi}= \frac{X(2)}{Y(2)} \,,
\end{equation} 
are related to the observables $M_\pi ,F_\pi$, to the LEC's $\bar{\ell}_3$
and $\bar{\ell}_4$ and to the NNLO remainders $\delta , \varepsilon$
by the following identities:
\begin{eqnarray}\label{identities} 
Y(2) &=& \frac{2(1-\delta)}{1-\varepsilon + [(1-\varepsilon)^2 - 2 \bar
  \ell_3 \xi (1-\delta)]^{(1/2)}}\,,\\
X(2) &=& 1 - \delta - (4 \bar{\ell}_4 - \bar{\ell}_3) \xi Y(2)^2/2\,,\\
Z(2) &=& 1 - \varepsilon - 2 \bar{\ell}_4 \xi Y(2) \,.
\end{eqnarray} 
In the last three equations, we have denoted
\begin{equation}\label{xi}
\xi = \frac{1}{16\pi^2} \frac{M^2_\pi}{F^2_\pi}\ .
\end{equation}       
The $\pi\pi$ subthreshold parameters $\alpha$ and $\beta$ can be expressed
similarly. Reading the LO and NLO perturbative contributions to
$F^2_\pi M^2_\pi \alpha$ and to $F^2_\pi \beta$ from Ref.~\cite{LET},
one obtains the identities:
\begin{eqnarray}\label{alpha}
\alpha &=& 1 - (1 + 3 \bar{\ell}_3 - 4 \bar{\ell}_4) \xi Y(2)^2 /2 +
\delta_\alpha\,, 
\\   \label{beta} 
\beta &=& 1 +  2 (\bar{\ell}_4 - 1) \xi Y(2) + \varepsilon_\beta\,.
\end{eqnarray}
It may be expected -- at least for the moment -- that the NNLO direct
remainders $\delta_\alpha$ and $\varepsilon_\beta$ are not more important
than the uncertainties in the determination of the parameters $\alpha$ and 
$\beta$ from experimental data.\\

\section{Determination of LEC's and order parameters}
\label{sec:l3l4}

\begin{figure}[ht]
\epsfxsize10cm
\centerline{\epsffile{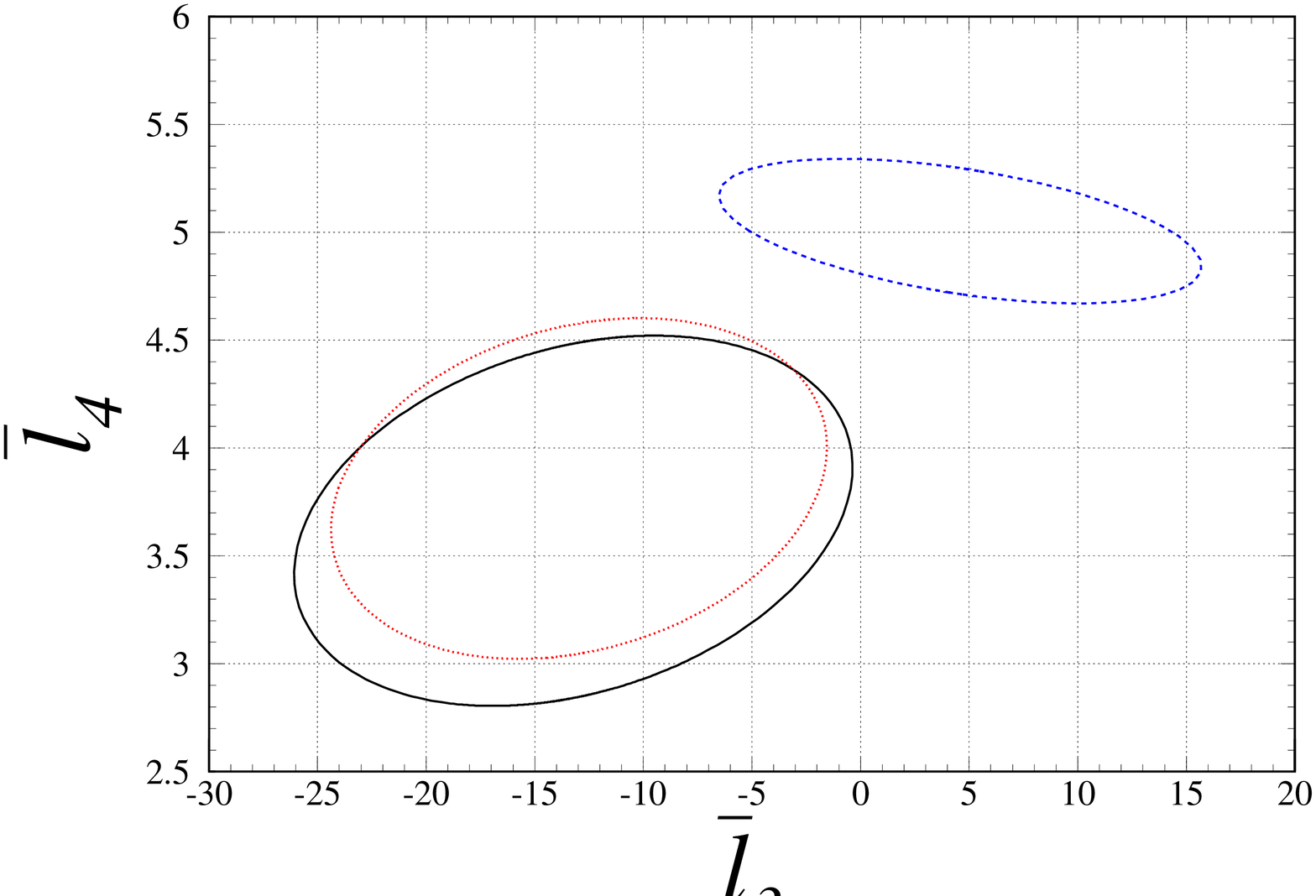}}

\vspace{0.2cm}

\caption{1-$\sigma$ ellipse in the
$\bar{\ell}_3 - \bar{\ell}_4$ plane, from the fit ``scalar'' (blue, dashed ellipse) and our fits (black, solid for the
``extended'' and red, dotted for the ``global''), using the linearized formulae.}
\label{fig:L3L4}
\end{figure}

If we expand the previous expressions of $\alpha$ and $\beta$ in
powers of $\xi$, we obtain the following (linearized) expressions in
term of the LEC's $\bar{\ell}_3$ and $\bar{\ell}_4$:
\begin{eqnarray}
\alpha-\beta &=& 3\xi(1-\bar{\ell}_3)/2\,, \label{alphalin} \\ 
\beta-1 &=& 2 \xi(\bar{\ell}_4-1)\,. \label{betalin}
\end{eqnarray}
This is an excellent approximation, unless $\bar{\ell}_3$ or
$\bar{\ell}_4$ become ``too'' large. However, even if one of them were
large, the non-linear equations (\ref{alpha}) and (\ref{beta}) of the
previous section would still be exact identities; moreover, the
definition of $\bar{\ell}_3$ and $\bar{\ell}_4$ in terms of two-point
functions is independent of their magnitudes.  We can use
Eqs.~(\ref{alphalin}) and (\ref{betalin}) to translate our
determination of $(\alpha,\beta)$ into a 1-$\sigma$ contour plot in
the $\bar{\ell}_3 - \bar{\ell}_4$ plane.  In Fig.~\ref{fig:L3L4}, we
show three ellipses corresponding to those in the $\alpha - \beta$
plane displayed in Fig.~\ref{fig:abcorr} above.

\begin{figure}[ht]
\epsfxsize10cm
\centerline{\epsffile{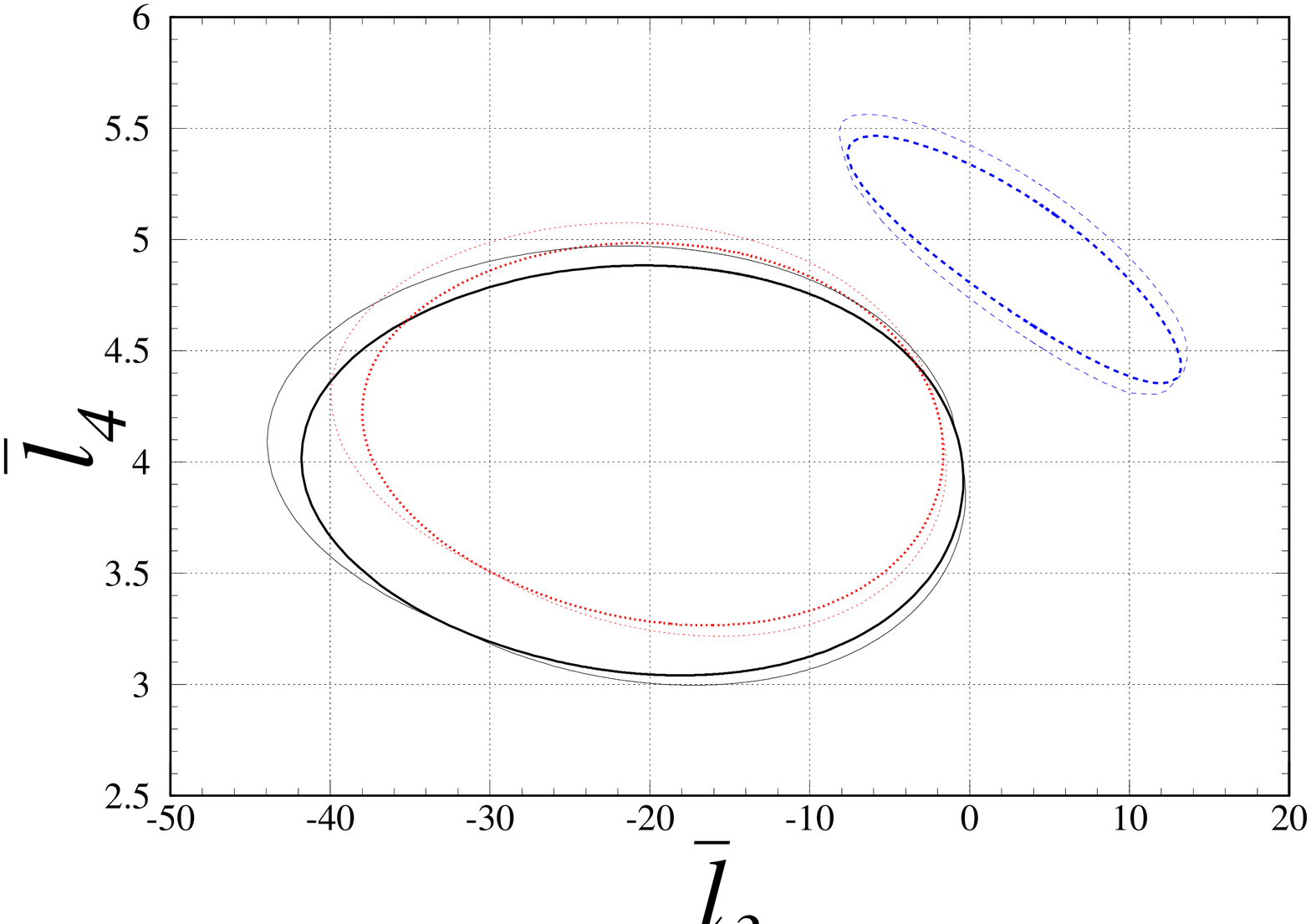}}

\vspace{0.2cm}

\caption{1-$\sigma$ ellipse in the $\bar{\ell}_3 - \bar{\ell}_4$
  plane, from the fit ``scalar'' (blue, dashed ellipse) and our
  fits (black, solid for the ``extended'' and red, dotted for the
``global''), using the non-linearized formulae. 
  The thinner lines indicate the domains allowed if
  $\varepsilon,\delta \leq 1\%$.}
\label{fig:L3L4nonpert}
\end{figure}

\begin{figure}[ht]
\epsfxsize10cm
\centerline{\epsffile{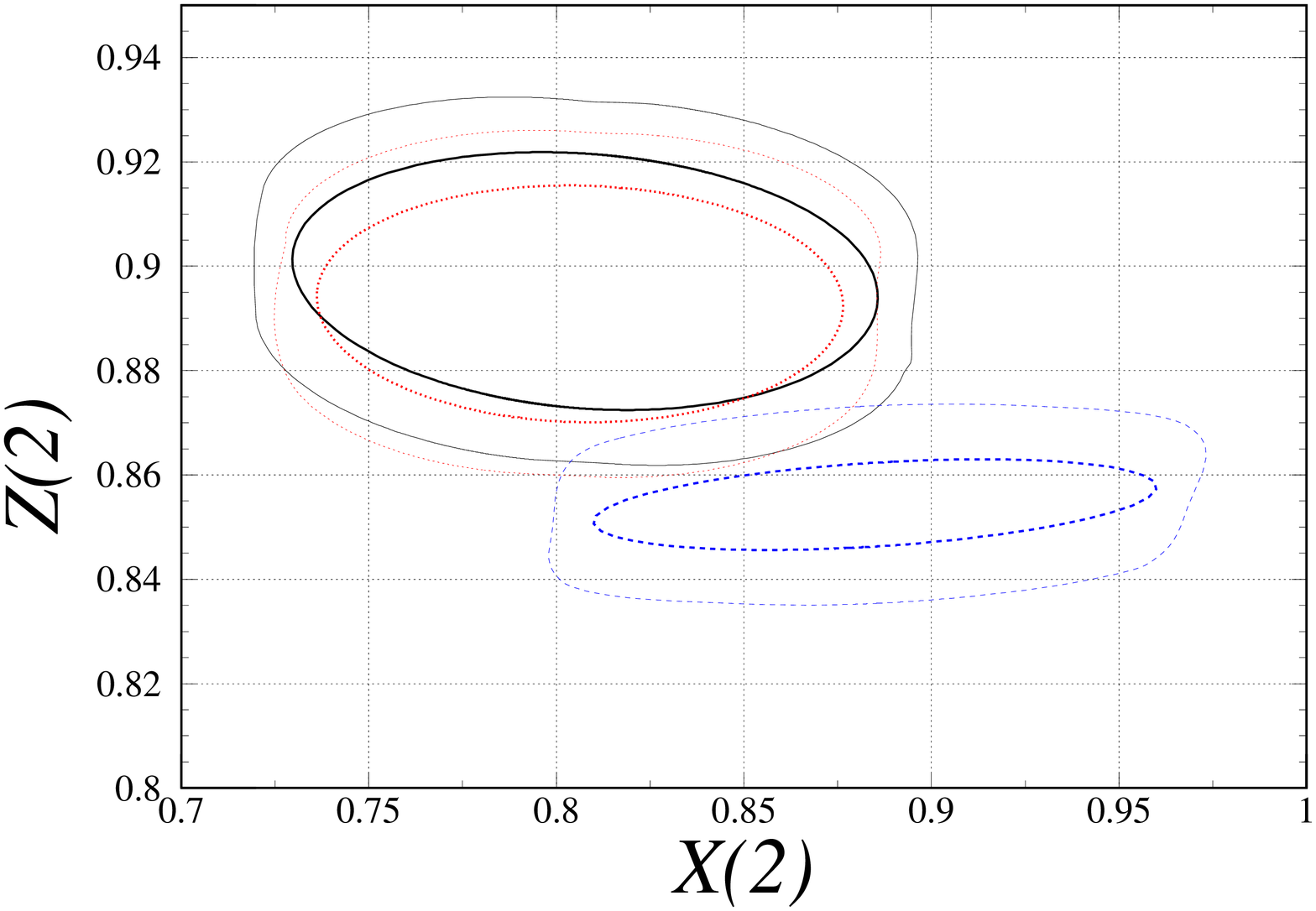}}
\caption{1-$\sigma$ ellipse in the
  $X(2) - Z(2) $ plane, from the fit ``scalar'' (blue, dashed 
  ellipse) and our fits (black, solid for the
``extended'' and red, dotted for the ``global'').  The thinner lines
  indicate the domains allowed if $\varepsilon,\delta \leq
  1\%$.} 
\label{fig:XZ}
\end{figure}

If we use the formulae Eqs.~(\ref{identities}), (\ref{alpha}) and
(\ref{beta}), but now {\em without linearizing}, the previous ellipses
are deformed, as shown in Fig.~\ref{fig:L3L4nonpert} (solid
lines). The corresponding contours in the $X(2) - Z(2) $ plane are
plotted in Fig.~\ref{fig:XZ}. Up to now, we have neglected the
indirect remainders $\delta$ and $\varepsilon$ as well as direct
remainders $\delta_{\alpha}$ and $\varepsilon_{\beta}$.  In the case
$N_f=2$, we expect these NNLO quantities to be less than 1\%, since
Eqs. (\ref{mass}) and (\ref{decayconstant}) are obtained by an
expansion in powers of the {\em nonstrange} quark mass.  This leads to
a (small) additional broadening of the 1-$\sigma$ regions, as seen in
the plots with thinner lines in Figs.~\ref{fig:L3L4nonpert} and
\ref{fig:XZ} ($\delta_{\alpha}$ and $\varepsilon_{\beta}$ are negligible
compared to the present uncertainty in the parameters $\alpha$ and
$\beta$).

It is clear from Fig.~\ref{fig:L3L4nonpert} that we obtain rather
large and negative values of $\bar{\ell_3}$, compared to the standard
expectation of 2.9 $\pm $ 2.4. This can be interpreted as a
consequence of an important OZI-rule violating transition
Eq.~(\ref{OZI}) leading to a larger value of the $N_f=3$, large-$N_c$
suppressed constant $L_6^r(M_{\rho})$, than usually assumed. Sum rule
estimates of the latter \cite{bachir,D} support this
interpretation. Notice that the OZI rule is an important ingredient of
the standard estimates of $\bar \ell_3$ \cite{GL1}. We will return to
this question elsewhere \cite{next}.

In Fig.~\ref{fig:XZ}, we see that the two-flavour GOR ratio $X(2)$ is
constrained (at one sigma): $X(2) = 0.81 \pm 0.09$.  In
Ref.~\cite{Girlanda}, this ratio has been theoretically analyzed,
including the $O(p^6)$ double chiral logarithms of Generalized
$\chi$PT, determining $X(2)$ as a function of $\alpha + 2 \beta$ and
other LEC's.  A combination of the results of the latter analysis with
the present (correlated) values for $\alpha$ and $\beta$ leads to a
range of values for $X(2)$ completely consistent with ours.  Further
examination of Fig.~\ref{fig:XZ} shows that $F/F_\pi$ is also limited
to a rather narrow band, $(F/F_\pi)^2 = 0.90 \pm 0.03$. Let us mention
that the difference seen in the $(\alpha,\beta)$ plane between the various 
fits reappears clearly here. The scalar fit (and the CGL prediction) 
favours larger values of $X(2)$ and lower values of $Z(2)$ than 
the global/extended fit.

\section{Comments on the correlation between scattering lengths and the
scalar radius of the pion}
\label{sec:scalar}

A relation between $2a_0^0-5a_0^2$ and the scalar radius of the pion
based on two-loop $\chi$PT has been derived in
Refs.~\cite{CGLPRL,CGLNPB}. For the current value of the scalar radius $\rad
= 0.61 \pm 0.04 \textrm{\ fm}^2$, this prediction results in the
narrow strip in the $a_0^0-a_0^2$ plane shown in Ref.~\cite{CGLPRL} and
reproduced here in Fig.~\ref{fig:fitres}. The accuracy of this
prediction is not only conditional on the experimental error, but also
on theoretical assumptions and ``rules'', which are a priori
reasonable and natural, but not necessarily true.

First of all, these assumptions concern the estimates of the $O(p^6)$
corrections~\cite{CGL1,CGLNPB}.  Although it is generally admitted
that for certain observables (e.g.  scattering lengths), these
corrections could be enhanced, it is expected that for quantities
receiving a weak contribution from $O(p^4)$ loops, the whole chiral
expansion will rapidly converge. An example of this rule of thumb
which is relevant for the present discussion is the quantity $C_1 -
M_\pi^2\rad/3$, which does not contain $O(p^4)$ logarithms at
all. ($C_1$ stands for a combination of subthreshold parameters
defined in Ref.~\cite{CGLNPB}.) We shall argue that this fact does not
prevent the corresponding $O(p^6)$ corrections from being relatively
important. (An alternative ``rule'' is conceivable~\cite{next}:
expand QCD correlation functions at kinematical points sufficiently
distant from all Goldstone boson singularities.  Such rule is a priori
not less or more natural.) Next, it is usually assumed that $O(p^6)$
counterterms at a suitable scale can be estimated via the narrow
resonance saturation~\cite{BCEG,CGLNPB,BCT}. In fact, already at
$O(p^4)$ this assumption fails in channels where $1/N_c$ --
corrections are large and/or the OZI-rule is strongly violated. This
is what likely happens in the scalar channel, which is particularly
relevant for the present discussion. Furthermore, the existing
resonance estimates of $O(p^6)$ counterterms have been so far based on
a ``resonance effective Lagrangian $\mathcal{L}_{\rm res}$'' involving
(and missing) the same resonances with the same ``minimal resonance
couplings'' as in Ref.~\cite{EGPR}, originally used to estimate the
$O(p^4)$ LEC's. It has often been argued~\cite{MS,KN} that additional
non-minimal couplings are necessary to avoid conflicts with the QCD
short-distance behaviour of two- and three-point functions, especially
if the latter involve (pseudo)scalar currents. The estimates of the
corresponding $O(p^6)$ counterterms $r_1 , r_2, r_3, r_4 , r_S$ can be
affected by these new resonance couplings. We shall return to the
resonance estimates of the $r_n$'s shortly.

\begin{figure}[ht]
\epsfxsize10cm
\centerline{\epsffile{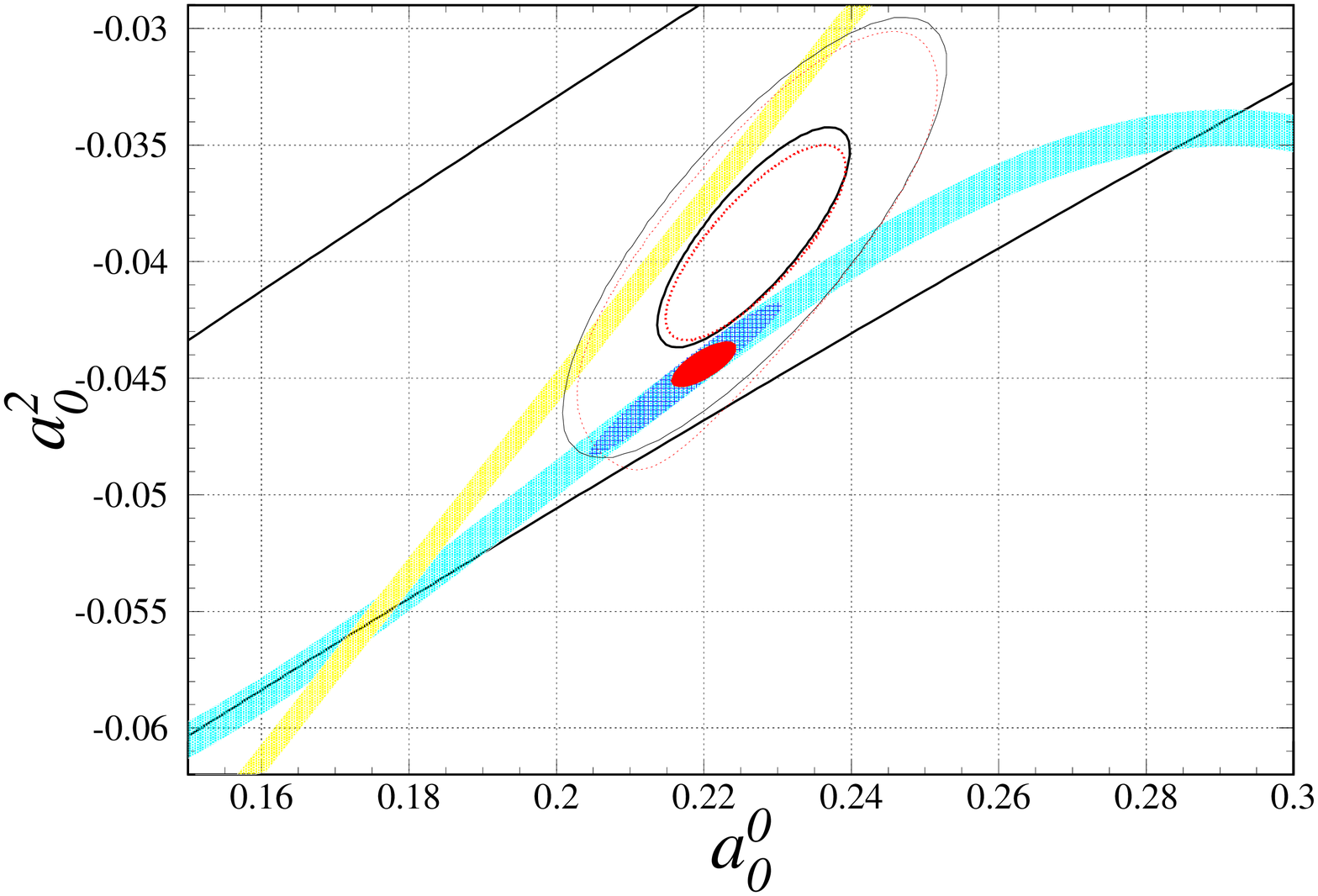}}
\caption{
Correlation Eq.~(\ref{eq:swave}) between $a_0^0$ and $a_0^2$, obtained from
the relation
between $2a_0^0-5a_0^2$ and the scalar radius of the pion. The shaded
oblique band (yellow)
corresponds to the $O(p^4)$ level [Eq.~(\ref{eq:swave}) with $\delta_a=0$],
using the dispersive estimate of Ref.~\cite{DGL}: $\rad=0.61\pm 0.04\
\textrm{fm}^2$.
The shaded curved strip (cyan)
represents the effect of $O(p^6)$ corrections $\delta_a$ according to the
prediction of Ref.~\cite{CGLPRL} using the same input value of $\langle r^2
\rangle_s$. The contours correspond to fits using either the Roy solutions of
Ref.~\cite{ACGL} (red, dotted ellipse -- ``global'') or our
parameterization of the Roy solutions (black, solid ellipse --
``extended''), with the errors of ACM(A) data enlarged as detailed in 
Appendix~\ref{sec:enlarged}. In each case, the thicker lines indicate the 1-$\sigma$
ellipse, and the thinner ones the 2-$\sigma$ contour. The other elements are
identical to Fig.~\ref{fig:fitres}.}
\label{fig:enlarged}
\end{figure}

Finally a remark should be made about the dispersive estimate of
$\rad$ in Ref.~\cite{DGL}, and the uncertainty related to it. The pion
scalar form-factor and radius are not experimentally measurable
quantities -- information about them can only come from indirect
theoretical constructions. In contrast to the case of the (observable)
vector form-factor, QCD does not restrict very much the high momentum
behaviour of the scalar form-factor. It even does not guarantee that
the latter satisfies an unsubtracted dispersion
relation. Consequently, the dispersive evaluation of $\rad$ suffers
from a certain model dependence concerning the higher momentum
contributions; this is usually not discussed in the literature. The
quoted uncertainty should not be interpreted outside the framework of
the model used in evaluating the scalar radius.

Most of these critical remarks are obviously not new. It might however
be useful to keep them in mind when discussing the origin of the
discrepancy between our model-free determination of scattering lengths
from the data and the CGL narrow strip prediction~\cite{CGLPRL}.  The
origin of the narrow strip Eq.~(\ref{eq:constscal}) is more easily
understood from Eq.~(3) of Ref.~\cite{CGLPRL}:
\begin{eqnarray} \label{eq:swave}
2a_0^0-5a_0^2&=&
  \frac{3M_\pi^2}{4\pi F_\pi^2}
    \left(1+\frac{1}{3}M_\pi^2\rad +\frac{41}{12}\xi\right)+\delta_a\\
&=&0.57158 + 0.05541 \left(\frac{\rad}{0.61\ \textrm{fm}^2}\right)+\delta_a
\, ,
\end{eqnarray}
where $\delta_a=O(m^3)$. It is worth stressing that the $O(p^6)$
contribution $\delta_a$ is essential for the numerical coherence of
Eq.~(\ref{eq:swave}). If the $O(p^6)$ contribution $\delta_a$ is dropped,
Eq.~(\ref{eq:swave}) reduces to the $O(p^4)$ low-energy theorem~\cite{LET}
relating $2a_0^0-5a_0^2$ and $\rad$. This model-independent $O(p^4)$ relation
is represented in Fig.~\ref{fig:enlarged} as a straight oblique band
corresponding to the estimate of the scalar radius
$\rad=0.61\pm 0.04\ \textrm{fm}^2$.

In order to reproduce the prediction of
Ref.~\cite{CGLNPB}: $2a^0_0-5a^2_0=0.663 \pm 0.006$, either the scalar radius should
be as large as  1.01 $\textrm{fm}^2$, or the $O(p^6)$ correction $\delta_a$
should move the $O(p^4)$ straight oblique band up to the curved strip along
the bottom of the Universal Band, reproduced in Fig.~\ref{fig:enlarged} from
Ref.~\cite{CGLPRL}. If we believe the estimate $\rad=0.61\pm 0.04\ \textrm{fm}^2$, 
and if Eq.~(\ref{eq:swave}) should hold inside the scalar or the
global ellipse, $\delta_a$ should take the values indicated in
Table~\ref{tab:deltaa}. The $O(p^6)$ correction amounts thus to 5\% in the scalar case,
and 3\% in the global one.

\begin{table}
\caption{Values of $\delta_a$ required for satisfying Eq.~(\ref{eq:swave}) and 
the estimate  $\rad = 0.61 \pm   0.04 \textrm{\ fm}^2$. The correlation
between $a_0^0$ and $a_0^2$ has been taken into account 
to compute the error on $2a_0^0-5a_0^2$.}\label{tab:deltaa}
\begin{tabular}{|c|rcl|rcl|}
\hline
 & \multicolumn{3}{c|}{Scalar}
 & \multicolumn{3}{c|}{Global} \\
\hline
 $a_0^0$         & 0.218   & $\pm  $ & 0.013  & 0.2279   &$\pm  $ & 0.012 \\
 $a_0^2$         & -0.0449 & $\pm  $ & 0.0033 & -0.0382 &$\pm  $ & 0.0038\\
 $2a_0^0-5a_0^2$ & 0.660   & $\pm  $ & 0.011  & 0.647   &$\pm  $ & 0.015\\
 $\delta_a$ & 0.033 & $\pm  $ & 0.012 & 0.020 & $\pm  $ & 0.015 \\
\hline
\end{tabular}
\end{table}

This size of $O(p^6)$ corrections is consistent with general expectations.
On one hand, NNLO contributions to ``smooth'' observables -- typically, QCD
correlation functions far from Goldstone boson singularities -- are
expected at 1\% level (see the NNLO remainders $\delta$ and $\varepsilon$ in
Sec.~\ref{sec:trunc}). On the other hand, we expect an enhancement of higher-order
corrections to infrared-singular ``threshold quantities'', such as scattering
lengths~\cite{CGL1,CGLNPB,KMSF2}. Let us emphasize that, in order to reproduce
the narrow strip of Ref.~\cite{CGLPRL}, the two-loop correction $\delta_a$
has to be known very precisely: this is no longer a matter of a
model-independent low-energy theorem \cite{LET} nor of an accurate knowledge
of $\rad$. In particular, the CGL prediction~\cite{CGLPRL} would be
brought into agreement with the present determination of $a_0^0$ and
$a_0^2$ by the ``global'' fit described in Sec.~\ref{sec:fits}, if the $O(p^6)$
correction $\delta_a$ were reduced by factor 2, even if the scalar
radius were not modified.

To understand the structure of the $O(p^6)$ correction $\delta_a$, we
follow Ref.~\cite{CGLNPB} and write:
\begin{equation}\label{eq:corr}
2a_0^0-5a_0^2 = \frac{3M_\pi^2}{4\pi F_\pi^2}C_1+M_\pi^4
\alpha_1+O(M_\pi^8) \,.
\end{equation}
$\alpha_1$ is a combination of phenomenological moments defined in Eq.~(6.4)
of Ref.~\cite{CGLNPB}. $C_1$ collects polynomial coefficients in the chiral
representation of the scattering amplitude.  Its expansion in powers of
$m$ reads~\cite{CGLNPB}:
\begin{equation}\label{eq:c1}
C_1 = 1+\frac{M_\pi^2}{3}\rad +\frac{23\xi}{420}+\xi^2\Delta_1+O(\xi^3)\,,
\end{equation}
where
\begin{eqnarray}\label{eq:D1}
\Delta_1& \equiv &
  \frac{-71\tilde{L}^2}{12}
  +\tilde{L}\Bigg\{-\frac{40}{9}\tilde{\ell}_1-\frac{80}{9}\tilde{\ell}_2 
-\frac{5}{2}\tilde{\ell}_3+4\tilde{\ell}_4+\frac{5393}{315}\Bigg\}\nonumber \\
&&
  -\tilde{\ell}_3\tilde{\ell}_4+\tilde{\ell}_4^2
  +\frac{1826}{315}\tilde{\ell}_1+\frac{3118}{315}\tilde{\ell}_2
  +\frac{79}{21}\tilde{\ell}_3   -\frac{144}{35}\tilde{\ell}_4
 \nonumber \\
&&  
  -\frac{521}{252}\pi^2+\frac{24221}{3024}
+\tilde{r}_2+4\tilde{r}_3-4\tilde{r}_4-2\tilde{r}_{S_2}\,,
\end{eqnarray}
contains the $O(p^4)$ constants
\begin{equation}\label{els}
\tilde{\ell_n}=\bar{\ell_n}-\tilde{L},\qquad
\tilde{L}=\log\frac{\mu^2}{M_\pi^2}\,,
\end{equation}
and the scale-dependent $O(p^6)$ counterterms
\begin{equation}\label{eq:rs}
\tilde{r}_n = (4 \pi)^4 r_n(\mu)\,.
\end{equation}

Comparing with Eq.~(\ref{eq:swave}), the $O(p^6)$ correction $\delta_a$ is
seen to consist of two parts:
\begin{equation}\label{decomposition}
\delta_a = \delta_M + \delta_1\,.
\end{equation}
$\delta_M$ is essentially the phenomenological moment $M_\pi^4
\alpha_1$ from which the leading infrared singularity has been
subtracted:
\begin{equation}\label{moment}
\delta_M = M_\pi^4 \alpha_1 - 4 \pi \frac{353}{35} \xi^2\,.
\end{equation}

With the help of Eq.~(6.8) of Ref.~\cite{CGLNPB}, one easily checks
that in the chiral limit $m \to 0$ all $O(p^4)$ contributions on the
right-hand side of Eq.~(\ref{moment}) exactly cancel. In practice,
however, $M_\pi^4 \alpha_1$ is obtained by evaluating infrared
singular sum rules using physical scattering lengths and not their
values in the chiral limit.  Consequently, the quantity $\delta_M$ is
expected to be sensitive to the infrared enhancement of higher order
corrections to the scattering lengths. One obtains:
\begin{equation}\label{alphascalar}
M_\pi^4 \alpha_1|_{\rm scalar} = 0.0604\pm   0.0054,
\quad \delta_M|_{\rm scalar}=0.034\pm 0.0054,
\end{equation}
\begin{equation}\label{alphaglobal}
M_\pi^4 \alpha_1|_{\rm global} = 0.0636\pm   0.005,
\quad \delta_M|_{\rm global}=0.037\pm 0.005,
\end{equation}
for $a_0^0$ and $a_0^2$ inside the ``scalar'' and ``global'' ellipses
respectively.
The second part of the $O(p^6)$ correction $\delta_a$ is due to $\Delta_1$:
\begin{equation}\label{secondpart}
\delta_1 = 12 \pi \xi^3 \Delta_1\,.
\end{equation}
As shown in Eq.~(\ref{eq:D1}), the latter depends on the $O(p^4)$
constants $\bar{\ell}_1,\bar{\ell}_2,\bar{\ell}_3,\bar{\ell}_4$ and on
the counterterms $\tilde{r}_2, \tilde{r}_3,\tilde{r}_4,
\tilde{r}_{S_2}$ describing $O(p^6)$ symmetry breaking effects in the
scalar sector.  The CGL prediction \cite{CGLPRL,CGLNPB} of the strong
correlation Eq.~(\ref{eq:constscal}) between scattering lengths originates
from a particular matching procedure worked out in Ref.~\cite{CGL1}, and
from the resonance estimate of $O(p^6)$ counterterms
\cite{BCEG,CGLNPB,BCT}, whose practical outcome is the fact that
$\delta_1$ is negligibly small and the whole two-loop correction
$\delta_a$ reduces to the contribution $\delta_M$ as shown in
Eq.~(\ref{alphascalar}).  This conclusion is independent of the actual
error in the dispersive estimate of the scalar radius $\rad = 0.61 \pm
0.04 {\textrm{\ fm}}^2 $.

The smallness of $\delta_1$ can indeed be justified within the
framework defined by the assumptions summarized at the beginning of
this section. To illustrate this point, let us concentrate first on
$a_0^0$ and $a_0^2$ inside the ellipse resulting from the ``scalar''
fit.  Choosing the scale $\mu = 770$ MeV ($\delta_1$ is scale-independent), we write
\begin{equation}\label{deltaone}
\delta_1 = \delta_{CT} + \delta_{\ell},
\end{equation}
where the counterterm part (at the scale $\mu$) is:
\begin{equation}\label{counterterm}
\delta_{CT} = 12 \pi \xi^3 ( \tilde{r}_2 + 4 \tilde{r}_3 - 4 \tilde{r}_4
                               - 2 \tilde{r}_{S2} )\,,
\end{equation}
whereas the loop part $\delta_{\ell}$ is given in terms of $O(p^4)$
constants $\bar{\ell}_1,\bar{\ell}_2,\bar{\ell}_3, \bar{\ell}_4$. We
take~\cite{CGLNPB}:
\begin{equation}\label{lonetwo}
\bar{\ell_1} = - 0.4 \pm   0.6,\qquad  \bar{\ell_2} = 4.3 \pm   0.1 \,,
\end{equation}
and for $\bar{\ell}_3,\bar{\ell}_4$ we use the result of the
``scalar'' fit represented by the corresponding ellipse in
Fig.~\ref{fig:L3L4}. This leads to the estimate $\delta_{\ell} =
0.0004 \pm 0.0083$. Using the resonance estimates of the
$r$'s~\cite{BCEG,BCT,CGLNPB} which are obtained according to the
prescription mentioned above and assuming that the typical mass scale
relevant in the scalar channel is $M_S = 1$ GeV, one obtains
$\delta_{CT} \sim 0.00025$. The authors of Ref.~\cite{CGLNPB} assume
that this estimate ($\tilde{r}_1 = -1.5, \tilde{r}_2 = 3.2, \tilde{r}_3
= -4.2, \tilde{r}_4 = -2.5, \tilde{r}_{S_2} = -0.7$) holds within a
factor of 2.  We can even relax this detailed estimate in favour of a
more crude order of magnitude:
\begin{equation}\label{estimate}
\tilde{r}_n = (4 \pi )^4 r_n(\mu) \sim \pm   \left(\frac{4\pi
F_\pi}{M_S}\right)^4 \,,
\end{equation}
which is expected to hold for $O(p^6)$ constants describing symmetry
breaking effects in the scalar channel. The crucial point here is the
dependence on the effective scalar mass $M_S$. As long as $M_S \sim 1$
GeV, $\delta_{CT}$ will remain small: adding individual $\tilde{r}_n$
contributions randomly ($\tilde{r}_n \sim \pm 1.8$), one gets in this
case $\delta_{CT} \sim \pm 0.0012$.  The whole two-loop correction
$\delta_a$ is then obtained by adding $\delta_M$ given by
Eq.~(\ref{alphascalar}) with the estimates of $\delta_{\ell}$ and
$\delta_{CT}$. One obtains $\delta_a = 0.034 \pm 0.010$ in agreement
with the value of $\delta_a$ required by the scalar fit and shown in
Table~\ref{tab:deltaa}.

We now return to the result of our paper indicating a deviation of the
CGL narrow strip from experiment. Let us assume for the moment that
the value of the scalar radius remains unchanged. Then, according to
Table~\ref{tab:deltaa}, the actual value of the two-loop correction
$\delta_a = 0.020 \pm 0.015$ is no longer saturated by $\delta_M =
0.037 \pm 0.005$ and consequently $\delta_1$ can no longer be
negligible. The loop part $\delta_{\ell}$ can be estimated as
before. Taking the same values Eq.~(\ref{lonetwo}) of $\ell_1 ,
\ell_2$ and using for $\ell_3 ,\ell_4$ the results of the global fit (see
the corresponding ellipse on Fig.~\ref{fig:L3L4}), one obtains in this
case\footnote{All the above estimates of the loop part of $\Delta_1$
can be reproduced by using its expression in terms of $a_0^0$ and
$a_0^2$, which arises from the matching with the Roy equations'
solutions.}  $\delta_{\ell} = 0.0096 \pm 0.0076$. This allows the
extraction of the required value of the $O(p^6)$ counterterm
combination $\delta_{CT}$. Parametrizing the possible variation of the
scalar radius by:
\begin{equation}\label{deltar}
\rad =0.61 \textrm{\ fm}^2 (1 + \delta_r) \,,
\end{equation}
one obtains the final estimate for the combined effect of
$\delta_{CT}$ and $\delta_r$ (our analysis does not allow us to
disentangle these corrections):
\begin{equation}\label{result}
\delta_{CT} + 0.05541\cdot\delta_r = - 0.0266 \pm 0.009\,.
\end{equation}
This means that both $O(p^6)$ contributions -- the subtracted infrared
singular moment $\delta_M$ and the symmetry breaking counterterms --
are basically of the same order of magnitude. This is still consistent
with the crude order of magnitude estimate of Eq.~(\ref{estimate}),
provided the effective mass scale $M_S$ characteristic of the scalar
channel contributions is reduced by a factor 2: $M_S \sim 500$
MeV. This could indeed be a rather natural (though rough and
qualitative) way how to account for the exceptional role of the
$\pi\pi$ continuum and of the OZI-rule violation in the scalar
channel. In this case, the estimate Eq.~(\ref{estimate}) leads to
$|\delta_{CT}| \sim  0.019$, in qualitative agreement with
Eq.~({\ref{result}), provided $\delta_{CT}$ is negative.

\section{Conclusion}
Low-energy $\pi\pi$ scattering has long been recognized as the golden
observable to access the chiral structure of QCD vacuum. From one side
the Roy equations allow a completely model-independent experimental
determination of the two S-wave scattering lengths $a_0^0$ and
$a_0^2$.  Chiral symmetry can then be used to translate this
information into knowledge of the LEC's of the chiral Lagrangian.  In
this paper we have followed these two steps using recent data on
$K_{e4}$ decays published by the E865 Collaboration.  Contrary to
previous analyses, we did not rely on any theoretical assumption about
the correlation between $a_0^0$ and $a_0^2$, but rather supplemented
the $K_{e4}$ data with existing data in the $I=2$ channel below
800~MeV. The result
\begin{equation}\label{eq:ours}
a_0^0 = 0.228 \pm 0.012, \quad a_0^2 = -0.0382 \pm 0.0038
\end{equation}
is compared with the theoretical relation between $2 a_0^0 - 5 a_0^2$
and the scalar radius of the pion \cite{CGLPRL} obtained in the
standard two-loop $\chi$PT. If the dispersive determination of the
latter, $\rad = (0.61 \pm 0.04) \textrm{\ fm}^2$ is used, one finds
a disagreement at the 1-$\sigma$ level.  It is possible to turn the
argument around, and interpret this discrepancy as a measurement of
$O(p^6)$ counterterms that contribute to the above theoretical
relations.  The latter come out larger than expected by the usual
resonance saturation assumptions; this fact might be the manifestation
of the exceptional status of the scalar channel, characterized by a
strong $\pi\pi$ continuum and OZI rule violation.

The same conclusions are reached once we include in the Roy equations
solutions the dependence on the phases at the matching point,
$\sqrt{s}= 800$~MeV, $\theta_0$ and $\theta_1$. These two quantities
are likely to represent the most important source of theoretical
uncertainty for the Roy solutions, aside from the one on the driving
terms. Our extended Roy solutions concretely parametrize the
theoretical error on the phaseshifts, for a given value of $a_0^0$ and
$a_0^2$.

Comparing the full two-loop standard $\chi$PT prediction for the two
scalar scattering lengths \cite{CGLNPB}, $a_0^0 = 0.220 \pm 0.005$ and
$a_0^2 = -0.0444 \pm 0.0010$ with Eq.~(\ref{eq:ours}), one finds agreement
for $a_0^0$, but a 1-$\sigma$ discrepancy persists for $a_0^2$.
This makes still more crucial the
outcome of new precise experiments on $\pi\pi$, which are either
planned \cite{NA48} or on-going \cite{dirac}. In particular, more
accurate $K_{e4}$ data in the region of higher $s_{\pi\pi}$ could
eventually allow a simultaneous determination of both $a_0^0$ and
$a_0^2$ from a single set of data on $\delta_0^0 - \delta_1^1$.  This
might be preferable to the procedure adopted in the present paper,
where E865 results are combined with the $\pi^+ \pi^+$-production data
of a different origin.   Our result on $a_0^0$ Eq.~(\ref{eq:ours})
agrees with one of the two determinations of the isoscalar scattering
length by the E865 collaboration \cite{pislak} that does not use the
narrow strip correlation between $a_0^0$ and $a_0^2$ as a theoretical
input \cite{CGLPRL}.  No determination of $a_0^2$ is reported in
\cite{pislak}.  In this respect it is interesting to notice
that the observed experimental correlation between $a_0^0$ and $a_0^2$
is positive and close to 1. The pionium lifetime experiments~\cite{dirac}
cannot distinguish $(a_0^0,a_0^2)$ pairs with the same $|a_0^0 - a_0^2|$,
in particular the global and scalar fits which lead to
$a^0_0-a_0^2=0.266 \pm 0.010$ and $0.263 \pm 0.010$ respectively.  

After determining the scattering lengths, we have studied the
consequences for the $\pi\pi$ subthreshold parameters, by means of a
matching procedure with the chiral amplitude. The influence of the
uncertainty in $\theta_0$ and $\theta_1$ is only apparent in the
parameters $\lambda_1$ and $\lambda_2$, whereas the others, in
particular $\alpha$ and $\beta$, are practically independent
thereof. These last two parameters are intimately related to the two
main order parameters of $SU(2)\times SU(2)$ chiral symmetry, the
quark condensate and the pion decay constant in the $SU(2)$ chiral
limit, keeping the strange quark mass at its physical value. The
result for the two-flavour GOR ratio $X(2)=0.81 \pm 0.09$ corresponds
to a large and negative central value for the LEC $\bar \ell_3\sim
-18$, to be compared with the standard expectation $\bar \ell_3 = 2.9
\pm 2.4$. Since this constant is rather sensitive to the OZI rule violating
constants $L_4$ and $L_6$ of the three-flavour chiral Lagrangian, this
conclusion seems to corroborate previous indications concerning the
size of OZI rule violation \cite{bachir,D} in the scalar channel,
namely the value of $L_6$.

Finally, we would like to stress that, although crucial for
understanding the pattern of $N_f=2$ chiral symmetry breaking,
$\pi\pi$ scattering alone does not tell us anything about the
important question of the dependence of chiral order parameters on
$N_f$. In particular it cannot be used to disentangle the ``genuine''
condensate of the purely massless theory, from the one ``induced'' by
massive (but light) strange quark pairs.  Nevertheless very accurate
low-energy data on $\pi\pi$ scattering still represent an essential
ingredient in a combined $SU(3) \times SU(3)$ analysis and
forthcoming determination \cite{next} of the three flavour condensate
$\Sigma(3)$, the quark mass ratios and other ``strange'' features of
the QCD vacuum.  For this program, new low-energy $\pi\pi$ and $\pi K$
scattering data are of obvious interest.\\

\begin{acknowledgement}
We would like to thank G. Colangelo, M. Knecht, B. Moussallam and H. 
Leutwyler for useful discussions and correspondence, J. Gasser 
for useful comments on the manuscript. Work partially supported by 
EEC-TMR contract ERBFMRXCT 98-0169 (EURODAPHNE). SD acknowledges 
partial support by PPARC, through grant PPA/G/S/ 1998/00530.  
LG acknowledges partial support from European Program HPRN-CT-2000-00149.
\end{acknowledgement}

\appendix

\section{Treatment of errors for ACM(A) data}\label{sec:enlarged}

As discussed at the end of Section~\ref{sec:fits},
it was proposed in Ref.~\cite{ACGL} to take into account (unknown) systematic errors
in method A of Ref.~\cite{hoogland}, by enlarging errors of ACM(A) 
solution in the following way. Consider for each energy bin the phase shift 
of solution A ($\delta_A \pm \Delta\delta_A$) and the one of solution B 
($\delta_B \pm \Delta\delta_B$). The difference
$|\delta_A-\delta_B|$ is supposed to reflect systematic errors of 
method A. The prescription advocated in Ref.~\cite{ACGL} consists then in
adding in quadrature this difference to the error $\Delta\delta_A$ quoted in
Ref.~\cite{hoogland}. The ACM(A) solution with enlarged errors is thus defined as:
\begin{equation}
\delta_A \pm \sqrt{(\Delta\delta_A)^2+(\delta_A-\delta_B)^2}
\end{equation}

The ``global'' and ``extended'' fits can be performed with these enlarged 
errors for ACM(A) phase shifts. The corresponding 1- and 2-$\sigma$ contours for the 
S-wave scattering lengths are plotted in Fig.~\ref{fig:enlarged}, while
Table~\ref{tab:enlarged} summarizes the changes occurring to the various 
quantities considered in this paper. Let us mention that
the nonlinear Eqs.~(\ref{identities}), (\ref{alpha}) and (\ref{beta})
 have been used to compute $\bar\ell_3$ and $\bar\ell_4$, 
and a contribution of the NNLO remainders $\delta,\varepsilon\leq 1\%$ is included 
in the error bars of $X(2)$ and $F^2/F_\pi^2$.

We see that enlarging errors for ACM(A) phase shifts slightly moves the 
global/extended contours in the $(a_0^0,a_0^2)$ plane towards the bottom of the 
Universal Band. The effect is however marginal: the 1-$\sigma$ discrepancy remains
between the scalar and the global/extended fits, and the derived quantities 
$\bar\ell_3$, $\bar\ell_4$, $X(2)$ and $F^2/F_\pi^2$ are almost unchanged.

\begin{table}
\caption{Results of the global and extended fits, considering for ACM(A)
phase shifts either the errors indicated in Ref.~\cite{hoogland}, or 
enlarged errors according to the prescription of Ref.~\cite{ACGL}.}
\label{tab:enlarged}
\begin{tabular}{|c|rcl|rcl|}
\hline
 & \multicolumn{6}{c|}{Global fit}\\
\hline
 & \multicolumn{3}{c|}{ACM(A) errors} & \multicolumn{3}{c|}{Enlarged errors}\\
\hline
$a_0^0$ & 0.228 & $\pm$ & 0.012 & 0.227 & $\pm$ & 0.012 \\
$a_0^2$ & -0.0382 & $\pm$ & 0.0038 & -0.0392 & $\pm$ & 0.0041 \\
$\chi^2$ & \multicolumn{3}{c|}{16.45} &  \multicolumn{3}{c|}{14.56} \\
\hline
$\alpha$ & 1.381 & $\pm$ & 0.242 & 1.334 & $\pm$ & 0.252\\
$\beta$  & 1.081 & $\pm$ & 0.023 & 1.088 & $\pm$ & 0.024\\ 
$\rho_{\alpha\beta}$ & \multicolumn{3}{c|}{-0.14} &
           \multicolumn{3}{c|}{-0.25}\\
\hline
$\bar\ell_3$ & -17.8 & $\pm$ & 15.3 & -14.1 & $\pm$ & 15.0 \\
$\bar\ell_4$ & 4.1 & $\pm$ & 0.9 & 4.3 & $\pm$ & 0.9 \\
$X(2)$ & 0.81 & $\pm$ & 0.07 & 0.82 & $\pm$ & 0.07 \\
$(F/F_\pi)^2$ & 0.89 & $\pm$ & 0.02 & 0.89 & $\pm$ & 0.03\\
\hline
\end{tabular}

\vspace{0.3cm}

\begin{tabular}{|c|rcl|rcl|}
\hline
 & \multicolumn{6}{c|}{Extended fit}\\
\hline
 & \multicolumn{3}{c|}{ACM(A) errors} & \multicolumn{3}{c|}{Enlarged errors} \\
\hline
$a_0^0$ & 0.228 & $\pm$ & 0.013 & 0.227 & $\pm$ & 0.013\\
$a_0^2$ & -0.0380 & $\pm$ & 0.0044 & -0.0389 & $\pm$ & 0.0047\\
$\chi^2$ & \multicolumn{3}{c|}{16.48} &  \multicolumn{3}{c|}{14.59}\\
\hline
$\alpha$ & 1.384 & $\pm$ & 0.267 & 1.340 & $\pm$ & 0.281\\
$\beta$  & 1.077 & $\pm$ & 0.025 & 1.084 & $\pm$ & 0.027\\
$\rho_{\alpha\beta}$ & \multicolumn{3}{c|}{-0.23} & 
	   \multicolumn{3}{c|}{-0.30}\\
\hline
  $\bar\ell_3$ & -18.5   & $\pm$ &  16.7  & -15.0 & $\pm$ & 16.3\\
$\bar\ell_4$ & 4.0 & $\pm$ & 0.9 & 4.2 & $\pm$ & 1.0\\
$X(2)$ & 0.81 & $\pm$ & 0.08 & 0.82 & $\pm$ & 0.08 \\
$(F/F_\pi)^2$ & 0.90 & $\pm$ & 0.03 & 0.89 & $\pm$ & 0.03\\
\hline
\end{tabular}

\vspace{0.3cm}

\end{table}

\section{Parametrization of the Roy solutions}\label{app:roy}

In Ref.~\cite{ACGL}, the solutions of Roy equations have been
described according to the Schenk parametrization
Eq.~(\ref{eq:deltaIl}). The Schenk parameters $X_\ell^I$
($X=A,B,C,D,s$) are functions of the scattering lengths, $a_0^0$ and
$a_0^2$, and the phase shifts at the matching point,
$\delta_0^0(s_0)\equiv\theta_0$ and
$\delta_1^1(s_0)\equiv\theta_1$. In Ref.~\cite{ACGL}, the Roy
equations were solved for the particular choice of phase shifts at the
matching point: $\theta_0=82.0^\circ$ and $\theta_1=108.9^\circ$, and
the dependence on $a_0^0$ and $a_0^2$ of the parameters $X=A,B,C,D,s$
was parametrized according to Eq.~(\ref{eq:paramACGL}).

We have followed the same procedure as in Ref.~\cite{ACGL}, with the
only difference that $\theta_0$ and $\theta_1$ are explicitly
treated as variables. After generating Roy solutions for
$\theta_0 \in \{ 78.9^\circ,$ $82.3^\circ,$ $85.7^\circ \} $ and
$\theta_1 \in \{106.9^\circ,$ $108.9^\circ,$ $110.9^\circ \} $, we then
performed a fit of the form of Eqs.~(\ref{eq:deltaIl}),
(\ref{eq:paramACGL}) and (\ref{eq:paramz}) with our solutions, in
order to obtain the parameters $a_i,b_i,c_i$ of Eq.~(\ref{eq:paramz}).
The Schenk parameters $s_0^0$, $s_1^1$ and $s_0^2$ are not
parametrized in the form of Eq.~(\ref{eq:paramACGL}), but are fixed by
the condition at the matching point $\delta^I_\ell(s_0)=\theta_I$.

The parameters $a_i$ are obtained by considering only the solution
$(\theta_0=82.3^\circ,\theta_1=108.9^\circ)$.  The parameters $b_i$
have then been obtained by fitting Roy solutions with
$\delta\theta_0\neq 0$ and $\delta\theta_1=0$, and the parameters
$c_i$ with $\delta\theta_0=0$ and $\delta\theta_1\neq 0$. At each
step, the fit has been performed at the level of the phase shifts, and
not of the Schenk parameters, in order to ensure a smooth dependence
on $(a_0^0,a_0^2,\theta_0,\theta_1)$.  We have checked that this
parametrization was adequate, even for solutions with both
non-vanishing $\delta\theta_0$ and $\delta\theta_1$.  The maximal gap
between any solution and our parametrization is at most 1\% in the
$I=0,1,2$ channels.

The coefficients resulting from the fit are expressed in units of 
$M_\pi$, and the phase shifts $\theta_{0,1,2}$ are in radians.
For $A_0^0$ and $A_0^2$, all the coefficients $a_i$, $b_i$, $c_i$ vanish,
apart from:
\begin{eqnarray}
A_0^0&:&a_1=a_2=0.225\,,\\
A_0^2&:&a_1=a_3=-0.03706\,,
\end{eqnarray}

$$
\begin{array}{|c | l || rcl | rcl | rcl|}
\hline
\textrm{Par.} & z_i & \multicolumn{3}{c|}{a_i} & 
\multicolumn{3}{c|}{b_i} & \multicolumn{3}{c|}{c_i}\\
\hline
 &  1 & .3617&\cdot & 10^{-1} & -.1716&\cdot & 10^{-2} & -.3860&\cdot & 10^{-2}\\
 &  2 & .1574&\cdot & 10^{-1} & -.2448&\cdot & 10^{-2} & -.3384&\cdot & 10^{-3}\\
 &  3 & .1057&\cdot & 10^{-1} & -.1774&\cdot & 10^{-2} & -.2510&\cdot & 10^{-4}\\
 &  4 & -.1782&\cdot & 10^{-2} & -.1025&\cdot & 10^{-1} & -.4312&\cdot & 10^{-2}\\
A_1^1 &  5 & .2572&\cdot & 10^{-3} & -.4649&\cdot & 10^{-2} & -.1705&\cdot & 10^{-2}\\
 &  6 & -.2872&\cdot & 10^{-3} & .1046&\cdot & 10^{-2} & -.3467&\cdot & 10^{-2}\\
 &  7 & .8311&\cdot & 10^{-2} & -.9152&\cdot & 10^{-2} & -.3637&\cdot & 10^{-2}\\
 &  8 & -.2603&\cdot & 10^{-2} & -.1489&\cdot & 10^{-1} & .2188&\cdot & 10^{-2}\\
 &  9 & .1247&\cdot & 10^{-2} & .7639&\cdot & 10^{-3} & -.1340&\cdot & 10^{-2}\\
 & 10 & -.1186&\cdot & 10^{-3} & .4371&\cdot & 10^{-2} & .1128&\cdot & 10^{-4}\\
\hline
\end{array}
$$
$$
\begin{array}{|c | l || rcl | rcl | rcl|}
\hline
\textrm{Par.} & z_i & \multicolumn{3}{c|}{a_i} & 
\multicolumn{3}{c|}{b_i} & \multicolumn{3}{c|}{c_i}\\
\hline
 &  1 & .2482 &&& .4902&\cdot & 10^{-1} & .1282&\cdot & 10^{-1}\\
 &  2 & .1997 &&& .1630 &&& -.3179&\cdot & 10^{-3}\\
 &  3 & .1285 &&& .1137 &&& .1640&\cdot & 10^{-3}\\
 &  4 & .1831&\cdot & 10^{-1} & -.1185 &&& .6305&\cdot & 10^{-1}\\
B_0^0 &  5 & .9970&\cdot & 10^{-2} & -.6395&\cdot & 10^{-2} & .1104&\cdot & 10^{-1}\\
 &  6 & .4846&\cdot & 10^{-1} & .3431 &&& -.1661&\cdot & 10^{-1}\\
 &  7 & -.3888&\cdot & 10^{-2} & -.1598&& & .4322&\cdot & 10^{-1}\\
 &  8 & -.8912&\cdot & 10^{-2} & .5183 &&& -.3067&\cdot & 10^{-1}\\
 &  9 & -.4265&\cdot & 10^{-2} & .4161&\cdot & 10^{-1} & .8623&\cdot & 10^{-2}\\
 & 10 & -.3232&\cdot & 10^{-2} & -.1073 &&& .2976&\cdot & 10^{-2}\\
\hline
\end{array}
$$
$$
\begin{array}{|c | l || rcl | rcl | rcl|}
\hline
\textrm{Par.} & z_i & \multicolumn{3}{c|}{a_i} & 
\multicolumn{3}{c|}{b_i} & \multicolumn{3}{c|}{c_i}\\
\hline
&  1 & .1135&\cdot & 10^{-3} & -.1685&\cdot & 10^{-3} & -.6043&\cdot & 10^{-3}\\
 &  2 & -.2094&\cdot & 10^{-2} & -.3429&\cdot & 10^{-3} & -.5583&\cdot & 10^{-4}\\
 &  3 & -.8626&\cdot & 10^{-3} & -.2467&\cdot & 10^{-3} & -.2205&\cdot & 10^{-4}\\
 &  4 & .2911&\cdot & 10^{-3} & -.8897&\cdot & 10^{-3} & -.5793&\cdot & 10^{-3}\\
B_1^1 &  5 & .7343&\cdot & 10^{-4} & -.4099&\cdot & 10^{-3} & -.2258&\cdot & 10^{-3}\\
 &  6 & .2063&\cdot & 10^{-3} & -.4832&\cdot & 10^{-3} & -.6376&\cdot & 10^{-3}\\
 &  7 & .5294&\cdot & 10^{-3} & -.6346&\cdot & 10^{-3} & -.3879&\cdot & 10^{-3}\\
 &  8 & -.3372&\cdot & 10^{-3} & -.2347&\cdot & 10^{-2} & .9292&\cdot & 10^{-5}\\
 &  9 & -.1564&\cdot & 10^{-3} & .1032&\cdot & 10^{-4} & -.1169&\cdot & 10^{-4}\\
 & 10 & -.1301&\cdot & 10^{-4} & .8137&\cdot & 10^{-3} & -.1051&\cdot & 10^{-3}\\
\hline
\end{array}
$$
$$
\begin{array}{|c | l || rcl | rcl | rcl|}
\hline
\textrm{Par.} & z_i & \multicolumn{3}{c|}{a_i} & 
\multicolumn{3}{c|}{b_i} & \multicolumn{3}{c|}{c_i}\\
\hline
&  1 & -.8567&\cdot & 10^{-1} & -.5496&\cdot & 10^{-2} & .1526&\cdot & 10^{-2}\\
 &  2 & -.1561&\cdot & 10^{-1} & .1510&\cdot & 10^{-2} & -.6254&\cdot & 10^{-3}\\
 &  3 & -.8722&\cdot & 10^{-2} & .9679&\cdot & 10^{-3} & .2538&\cdot & 10^{-3}\\
 &  4 & .9872&\cdot & 10^{-2} & .1001&\cdot & 10^{-1} & .2140&\cdot & 10^{-1}\\
B_0^2 &  5 & .2176&\cdot & 10^{-1} & .3724&\cdot & 10^{-2} & .3595&\cdot & 10^{-2}\\
 &  6 & .3338&\cdot & 10^{-1} & -.1050&\cdot & 10^{-1} & -.5945&\cdot & 10^{-2}\\
 &  7 & -.2051&\cdot & 10^{-1} & .4012&\cdot & 10^{-1} & .1157&\cdot & 10^{-1}\\
 &  8 & -.5171&\cdot & 10^{-1} & .7078&\cdot & 10^{-2} & .1593&\cdot & 10^{-2}\\
 &  9 & -.5929&\cdot & 10^{-1} & -.6046&\cdot & 10^{-2} & .1382&\cdot & 10^{-2}\\
 & 10 & -.2247&\cdot & 10^{-1} & .4017&\cdot & 10^{-2} & -.1490&\cdot & 10^{-2}\\
\hline
\end{array}
$$
$$
\begin{array}{|c | l || rcl | rcl | rcl|}
\hline
\textrm{Par.} & z_i & \multicolumn{3}{c|}{a_i} & 
\multicolumn{3}{c|}{b_i} & \multicolumn{3}{c|}{c_i}\\
\hline
 &  1 & -.1652&\cdot & 10^{-1} & .2246&\cdot & 10^{-1} & .3320&\cdot & 10^{-2}\\
 &  2 & .3280&\cdot & 10^{-2} & .5387&\cdot & 10^{-1} & .9391&\cdot & 10^{-4}\\
 &  3 & .1127&\cdot & 10^{-1} & .2911&\cdot & 10^{-1} & .2303&\cdot & 10^{-3}\\
 &  4 & .1367&\cdot & 10^{-1} & .1198 & & & .9361&\cdot & 10^{-2}\\
C_0^0 &  5 & .1606&\cdot & 10^{-1} & .5107&\cdot & 10^{-1} & .1440&\cdot & 10^{-3}\\
 &  6 & .2990&\cdot & 10^{-1} & -.1170&\cdot & 10^{-1} & .1345&\cdot & 10^{-2}\\
 &  7 & -.5982&\cdot & 10^{-2} & .9021&\cdot & 10^{-1} & .1428&\cdot & 10^{-1}\\
 &  8 & .1923&\cdot & 10^{-2} & .9601&\cdot & 10^{-1} & -.4036&\cdot & 10^{-2}\\
 &  9 & .1106&\cdot & 10^{-1} & .2148&\cdot & 10^{-1} & -.1501&\cdot & 10^{-2}\\
 & 10 & .3809&\cdot & 10^{-2} & -.2854&\cdot & 10^{-1} & .2780&\cdot & 10^{-2}\\
\hline
\end{array}
$$
$$
\begin{array}{|c | l || rcl | rcl | rcl|}
\hline
\textrm{Par.} & z_i & \multicolumn{3}{c|}{a_i} & 
\multicolumn{3}{c|}{b_i} & \multicolumn{3}{c|}{c_i}\\
\hline
&  1 & -.7257&\cdot & 10^{-4} & -.1076&\cdot & 10^{-4} & -.8750&\cdot & 10^{-4}\\
 &  2 & .2234&\cdot & 10^{-3} & -.4577&\cdot & 10^{-4} & -.8053&\cdot & 10^{-5}\\
 &  3 & .3718&\cdot & 10^{-4} & -.3531&\cdot & 10^{-4} & -.6497&\cdot & 10^{-5}\\
 &  4 & .2259&\cdot & 10^{-4} & .2031&\cdot & 10^{-4} & -.7306&\cdot & 10^{-4}\\
C_1^1 &  5 & .1216&\cdot & 10^{-4} & -.2042&\cdot & 10^{-4} & -.2856&\cdot & 10^{-4}\\
 &  6 & .4075&\cdot & 10^{-4} & -.1625&\cdot & 10^{-3} & -.1121&\cdot & 10^{-3}\\
 &  7 & -.1238&\cdot & 10^{-4} & -.3676&\cdot & 10^{-4} & -.2568&\cdot & 10^{-4}\\
 &  8 & .1103&\cdot & 10^{-3} & -.3679&\cdot & 10^{-3} & -.5010&\cdot & 10^{-4}\\
 &  9 & .3813&\cdot & 10^{-4} & -.5706&\cdot & 10^{-5} & .3202&\cdot & 10^{-4}\\
 & 10 & .3531&\cdot & 10^{-4} & .1373&\cdot & 10^{-3} & -.3439&\cdot & 10^{-4}\\
\hline
\end{array}
$$
$$
\begin{array}{|c | l || rcl | rcl | rcl|}
\hline
\textrm{Par.} & z_i & \multicolumn{3}{c|}{a_i} & 
\multicolumn{3}{c|}{b_i} & \multicolumn{3}{c|}{c_i}\\
\hline
&  1 & -.7557&\cdot & 10^{-2} & .2648&\cdot & 10^{-2} & -.5166&\cdot & 10^{-3}\\
 &  2 & .3425&\cdot & 10^{-1} & -.2038&\cdot & 10^{-2} & .5412&\cdot & 10^{-3}\\
 &  3 & .2830&\cdot & 10^{-1} & -.9686&\cdot & 10^{-3} & .2995&\cdot & 10^{-3}\\
 &  4 & .3342&\cdot & 10^{-2} & .5536&\cdot & 10^{-2} & -.5538&\cdot & 10^{-2}\\
C_0^2 &  5 & .1391&\cdot & 10^{-1} & .7956&\cdot & 10^{-3} & -.2012&\cdot & 10^{-2}\\
 &  6 & .2375&\cdot & 10^{-1} & .1775&\cdot & 10^{-2} & .2675&\cdot & 10^{-2}\\
 &  7 & -.3024&\cdot & 10^{-1} & -.1924&\cdot & 10^{-1} & .6680&\cdot & 10^{-4}\\
 &  8 & -.9323&\cdot & 10^{-1} & -.2108&\cdot & 10^{-2} & .2173&\cdot & 10^{-2}\\
 &  9 & -.8813&\cdot & 10^{-1} & .4251&\cdot & 10^{-2} & -.2462&\cdot & 10^{-2}\\
 & 10 & -.2679&\cdot & 10^{-1} & -.3504&\cdot & 10^{-2} & .1984&\cdot & 10^{-2}\\
\hline
\end{array}
$$
$$
\begin{array}{|c | l || rcl | rcl | rcl|}
\hline
\textrm{Par.} & z_i & \multicolumn{3}{c|}{a_i} & 
\multicolumn{3}{c|}{b_i} & \multicolumn{3}{c|}{c_i}\\
\hline 
 &  1 & -.6396&\cdot & 10^{-3} & .7978&\cdot & 10^{-3} & .6667&\cdot & 10^{-3}\\
 &  2 & -.4143&\cdot & 10^{-2} & .5649&\cdot & 10^{-2} & -.5508&\cdot & 10^{-4}\\
 &  3 & -.3708&\cdot & 10^{-2} & .5227&\cdot & 10^{-2} & .1462&\cdot & 10^{-3}\\
 &  4 & -.4016&\cdot & 10^{-2} & -.6414&\cdot & 10^{-2} & -.8673&\cdot & 10^{-3}\\
D_0^0 &  5 & -.3159&\cdot & 10^{-2} & -.3022&\cdot & 10^{-2} & -.9427&\cdot & 10^{-3}\\
 &  6 & -.7352&\cdot & 10^{-2} & .1584&\cdot & 10^{-1} & .2274&\cdot & 10^{-2}\\
 &  7 & -.1305&\cdot & 10^{-2} & -.1363&\cdot & 10^{-1} & .3488&\cdot & 10^{-2}\\
 &  8 & -.4523&\cdot & 10^{-2} & .1960&\cdot & 10^{-1} & .1146&\cdot & 10^{-2}\\
 &  9 & -.4581&\cdot & 10^{-2} & -.2917&\cdot & 10^{-3} & -.1778&\cdot & 10^{-2}\\
 & 10 & -.1272&\cdot & 10^{-2} & -.4082&\cdot & 10^{-2} & .1184&\cdot & 10^{-2}\\
\hline
\end{array}
$$
$$
\begin{array}{|c | l || rcl | rcl | rcl|}
\hline
\textrm{Par.} & z_i & \multicolumn{3}{c|}{a_i} & 
\multicolumn{3}{c|}{b_i} & \multicolumn{3}{c|}{c_i}\\
\hline
&  1 & .6607&\cdot & 10^{-7} & -.1767&\cdot & 10^{-6} & -.1271&\cdot & 10^{-4}\\
 &  2 & -.1750&\cdot & 10^{-4} & -.5895&\cdot & 10^{-5} & -.8847&\cdot & 10^{-6}\\
 &  3 & -.6507&\cdot & 10^{-5} & -.5144&\cdot & 10^{-5} & -.1517&\cdot & 10^{-5}\\
 &  4 & -.3851&\cdot & 10^{-5} & .1657&\cdot & 10^{-4} & -.7559&\cdot & 10^{-5}\\
D_1^1 &  5 & .4987&\cdot & 10^{-6} & .2201&\cdot & 10^{-5} & -.3089&\cdot & 10^{-5}\\
 &  6 & .1953&\cdot & 10^{-6} & -.3159&\cdot & 10^{-4} & -.1827&\cdot & 10^{-4}\\
 &  7 & -.2797&\cdot & 10^{-4} & .3893&\cdot & 10^{-5} & .1194&\cdot & 10^{-5}\\
 &  8 & .1604&\cdot & 10^{-4} & -.5762&\cdot & 10^{-4} & -.1570&\cdot & 10^{-4}\\
 &  9 & -.1183&\cdot & 10^{-4} & -.9919&\cdot & 10^{-6} & .9930&\cdot & 10^{-5}\\
 & 10 & -.7835&\cdot & 10^{-5} & .2179&\cdot & 10^{-4} & -.7949&\cdot & 10^{-5}\\
\hline
\end{array}
$$
$$
\begin{array}{|c | l || rcl | rcl | rcl|}
\hline
\textrm{Par.} & z_i & \multicolumn{3}{c|}{a_i} & 
\multicolumn{3}{c|}{b_i} & \multicolumn{3}{c|}{c_i}\\
\hline
&  1 & .1980&\cdot & 10^{-3} & .1510&\cdot & 10^{-3} & -.2527&\cdot & 10^{-4}\\
 &  2 & -.2572&\cdot & 10^{-2} & -.5907&\cdot & 10^{-4} & .1149&\cdot & 10^{-4}\\
 &  3 & -.2024&\cdot & 10^{-2} & -.2137&\cdot & 10^{-4} & .1067&\cdot & 10^{-4}\\
 &  4 & .1600&\cdot & 10^{-2} & .5689&\cdot & 10^{-3} & -.2189&\cdot & 10^{-3}\\
D_0^2 &  5 & .1790&\cdot & 10^{-3} & .1280&\cdot & 10^{-3} & -.7452&\cdot & 10^{-4}\\
 &  6 & .1228&\cdot & 10^{-2} & -.3551&\cdot & 10^{-4} & .9342&\cdot & 10^{-4}\\
 &  7 & .9168&\cdot & 10^{-3} & -.7961&\cdot & 10^{-3} & -.1405&\cdot & 10^{-3}\\
 &  8 & .4960&\cdot & 10^{-2} & -.1981&\cdot & 10^{-3} & .7174&\cdot & 10^{-4}\\
 &  9 & .5225&\cdot & 10^{-2} & .2966&\cdot & 10^{-3} & -.8173&\cdot & 10^{-4}\\
 & 10 & .1550&\cdot & 10^{-2} & -.1694&\cdot & 10^{-3} & .6938&\cdot & 10^{-4}\\
\hline
\end{array}
$$
We determine 
$s_0^0$, $s_1^1$ and $s_0^2$ by the conditions:
\begin{equation}
\delta_0^0(s_0)\equiv\theta_0\ , \qquad
\delta_1^1(s_0)\equiv\theta_1\ , \qquad \delta_0^2(s_0)\equiv\theta_2\
\end{equation}
where $\theta_2(a_0^0,a_0^2,\theta_0,\theta_1)$ is parametrized following
Eqs.~(\ref{eq:paramACGL}) and (\ref{eq:paramz}), with the coefficients:

$$
\begin{array}{|c | l || rcl | rcl | rcl|}
\hline
\textrm{Par.} & z_i & \multicolumn{3}{c|}{a_i} & 
\multicolumn{3}{c|}{b_i} & \multicolumn{3}{c|}{c_i}\\
\hline
 &  1 & -.3160 &&& .7038&\cdot & 10^{-1} & -.2480&\cdot & 10^{-1}\\
 &  2 & -.2355 &&& .2380&\cdot & 10^{-1} & .6701&\cdot & 10^{-2}\\
 &  3 & -.2021 &&& .1687&\cdot & 10^{-1} & .5869&\cdot & 10^{-2}\\
 &  4 & .4885&\cdot & 10^{-1} & .6057&\cdot & 10^{-1} & -.2094&\cdot & 10^{-1}\\
\theta_2 &  5 & -.1106&\cdot & 10^{-1} & .2317&\cdot & 10^{-1} & -.1128&\cdot & 10^{-1}\\
 &  6 & .8406&\cdot & 10^{-2} & .7702&\cdot & 10^{-1} & -.2254&\cdot & 10^{-1}\\
 &  7 & .3569&\cdot & 10^{-2} & .1531 &&& .1103&&\\
 &  8 & .3021&\cdot & 10^{-1} & .1027&\cdot & 10^{-2} & -.4945&\cdot & 10^{-2}\\
 &  9 & .2762&\cdot & 10^{-1} & .2859&\cdot & 10^{-2} & -.1297&\cdot & 10^{-1}\\
 & 10 & .7229&\cdot & 10^{-2} & .1513&\cdot & 10^{-1} & .1340&\cdot & 10^{-1}\\
\hline
\end{array}
$$

We have introduced the function
$\theta_2(a_0^0,a_0^2,\theta_0,\theta_1)$ to improve the accuracy on
$s_0^2$. The dependence of $\theta_2$ on the S-wave scattering lengths
and on the phases at the matching point is smoother than the analogous
dependence of $s^2_0$. A direct fit of $s^2_0$ using the form of
Eqs.~(\ref{eq:paramACGL}) and (\ref{eq:paramz}) would therefore induce
a loss of accuracy, compared to the procedure we follow here.

\end{document}